\documentclass[psfig]{mn2e}
\usepackage{graphicx}
\begin{document}
\title[Detecting $z>8$ QSOs by Near-IR Spectroscopy]
 {Detecting the Highest Redshift ({\bf $z>8$}) QSOs in a Wide,
Near Infrared Slitless Spectroscopic Survey}
\author[N. Roche, P.Franzetti, B.Garilli, G.Zamorani, A. Cimatti, E.Rossetti]
 {Nathan Roche$^1$, Paolo Franzetti$^2$, Bianca Garilli$^2$, Giovanni Zamorani$^1$,\\\\
{\rm \LARGE Andrea Cimatti$^3$, Emanuel Rossetti$^3$}\\
\thanks{nathanroche@mac.com; paolo@lambrate.inaf.it;
bianca@lambrate.inaf.it; giovanni.zamorani@oabo.inaf.it;
 a.cimatti@unibo.it: emanuel.rossetti@unibo.it}\\
$^1$ INAF--Osservatorio Astronomico di Bologna, Via Ranzani 1, Bologna 40127,
Italy.
$^2$ INAF--IASF, Via Bassini 15, I-20133, Milano, Italy.\\
$^3$ Dipartmento di Astronomia--Universita di Bologna,  Via Ranzani 1, Bologna 40127,
Italy.}

\bibliographystyle{unsrt} \bibliographystyle{unsrt}

\date{9 December 2011}

\pagerange{\pageref{firstpage}--\pageref{lastpage}} \pubyear{2011}

\maketitle

\label{firstpage}

\begin{abstract} 
We investigate the prospects of extending observations of high redshift QSOs from the current $z\sim 7$ to $z>8$ by means of a very wide-area near-infrared slitless spectroscopic survey, considering as an example the planned survey with the European Space Agency's {\it Euclid} telescope (scheduled for a 2019 launch). For any QSOs at $z>8.06$ the  strong Lyman-$\alpha$ line will enter the wavelength range of the {\it Euclid} near-infrared spectrometer and imaging photometer (NISP). We perform a detailed simulation of {\it Euclid}-NISP slitless spectroscopy (with the parameters of the wide survey) in an artificial field containing QSO spectra at all redshifts up to $z=12$ and to a faint limit $H=22.5$. QSO spectra are represented with a template based on an SDSS composite  spectrum, with the added effects of absorption from neutral hydrogen in the intergalactic medium. 

The spectra extracted from the simulation are analysed with an automated redshift finder, and a detection rate estimated as a function of $H$ magnitude and redshift (defined as the proportion of spectra with both correct redshift measurements and classifications). We show that, as expected, spectroscopic identification of  QSOs would reach deeper limits for the redshift ranges where either $\rm H\alpha$ ($0.67<z<2.05$) or Lyman-$\alpha$ ($z>8.06$) is visible. Furthermore, if photometrically-selected $z>8$ spectra can be re-examined and re-fitted to minimize the effects of spectral contamination, the QSO detection rate in the Lyman-$\alpha$ window will be increased by an estimated $\sim 60\%$ and will then be better here than at any other redshift, with an effective limit $H\simeq 21.5$. 

With an extrapolated rate of QSO evolution, we predict the {\it Euclid} wide (15000 $\rm deg^2$) spectroscopic survey will identify and measure spectroscopic redshifts for a total of 20--35 QSOs at $z>8.06$ (reduced slightly to 19--33 if we apply a small correction for missed weak-lined QSOs). However, for a model with a faster rate of evolution, this prediction goes down to 4 or 5. 
In any event, the survey will give important constraints on the evolution of QSO at $z>8$ and therefore the formation of the first super-massive black holes.
The $z>8.06$ detections would be very luminous objects (with $M_B=-26$ to -28) and many would also be detectable by the proposed {\it Wide Field X-ray Telescope}.
\end{abstract}


\section{Introduction} Luminous QSOs, powered by supermassive black holes (SMBH) of the order of $\sim 10^9$ Solar masses ($M_{\odot}$) have been spectroscopically identified  at $z\sim 6$--7, 
when the age of the Universe was $<1$ Gyr, but their even earlier evolution and (necessarily rapid) formation is as yet unobserved. 

The luminosity density of active galactic nuclei (AGN)  
increases strongly with redshift from $z=0$ to $z\sim 2$, but beyond $z\sim 3$ the comoving number density is seen to decline exponentially,  approximately as $\Phi^*\propto e^{-z}$ in both optical and X-ray surveys (Schmidt, Schneider and Gunn 1995, Fontanot et al. 2007a, Brusa et al. 2009). Even so, in the optical-spectroscopic Sloan Digital Sky Survey (SDSS), some QSOs have been identified to $z\simeq 6$ (Fan et al. 2003, 2004, 2006) with relatively bright apparent magnitudes of $z_{AB}\sim 20$, which at these redshifts correspond to extremely high optical-UV luminosities of $M_B=-27$ to -28. At fainter magnitudes ($z_{AB}\simeq 21$--22) more $z\sim 6$ QSOs have been identified in the SDSS deep strip (Jiang et al. 2008) and the Canada-France High Redshift Quasar Survey (CFHQZ; Willott et al.  2010), in most cases with strong Lyman-$\alpha$ emission lines in their spectra, bringing the total of confirmed QSOs at $5.74<z<6.43$ to at least 40.

 Willott et al. (2010b) found another QSO (CFHQS J021013-045620) at $z=6.44$, setting a new distance record. This was again surpassed when Mortlock et al. (2011) reported the discovery in the UKIRT Infrared Deep Sky Survey (UKIDSS) of a very  luminous and  massive ($M_{1450} = -26.6$, $M_{BH}\simeq 2\times 10^9 M_{\odot}$) QSO (ULAS J1120+0641) at $z=7.08$ (at an age of the Universe 0.77 Gyr). Its SMBH is accreting at approximately the Eddington-limit rate, which for an efficiency $\epsilon=0.1$ implies an e-folding time for SMBH growth of 45 Myr (Willott et al. 2011b). Clearly, to have grown up from a seed black hole of stellar ($\leq 100 M_{\odot}$) or star cluster mass ($\sim 10^4 M_{\odot}$, possibly even more) at $\sim 1$ dex/0.1 Gyr, it must have formed very early and accreted at a maximal rate. The existence of such a massive and luminous QSO at $z\sim 7$  is already a challenge for formation models; yet, especially if seed black holes can be very massive, there could be similar QSOs awaiting discovery at even higher redshifts.

Models for QSO number counts at $z>6$ may differ widely. For example, in the models of  Fontanot et al. (2007b), if the observed exponential negative density  evolution (as ${d\Phi^*}\over{dz}$) at $3<z<6$ can simply be extrapolated to  $z>6$, it is likely that in the near future some QSOs will be found at $z\geq 8$. However, if all QSOs at $z\geq 6$ are accreting at the maximum possible rate (Eddington luminosity and maximum duty cycle of 1), this implies a much faster rate of evolution, so that QSOs completely disappear from detectability at $z\sim 7.5$.

 More recently, Shankar et al. (2010) interpreted the strong clustering of $z\sim 3$--4 QSOs as evidence they reside in extremely massive halos and further that the duty cycles are high and increase to approach unity at $z=6$. As such massive halos would become very rare at $z>6$ and the duty cycle could not further increase, they predict that the negative evolution of the QSO density (${d\Phi^*}\over{dz}$) would steepen markedly (by a factor $\sim 2$) at $z\geq 
6$. Again this greatly reduces the detectable numbers at even higher redshifts relative to the extrapolated model. However, this is not a certainty; the statistical errors on the clustering ($r_0$) are large and the interpretation (in terms of halo mass, duty cycle etc.) is model dependent.

Trakhtenbrot et al. (2011), by measuring MgII line widths for $z\simeq 4.8$ QSOs, find their SMBH masses to average almost $10^9M_{\odot}$, with high (median 0.59) but widely spread Eddington ratios. All these SMBH must have at times accreted very rapidly to have grown to these masses 
 within the $\leq 1$ Gyr available. Yet the moderate evolution of SMBH masses over $2.4<z<4.8$ could only be fitted by assuming low duty cycles of 0.1--0.2. 
If some fraction of AGN had low duty cycles even at higher redshifts, their epoch of very rapid mass growth would be pushed back earlier.

There are QSOs with the same high SMBH masses, but in smaller numbers, at $6<z<7$. A scenario of very rapid ($\sim$ Eddington-limit) but intermittent (merger-driven?) accretion, with diverse black hole growth histories, would again raise the possibility of finding some very luminous QSOs out at $z>8$. However, as the e-folding time for growth is constrained by the Eddington limit, the maximum redshift for detectable (very luminous) QSOs may set lower limits on the masses of the seed black holes (Willott 2011). Finding luminous QSOs at $z>8$ or even $z\sim 10$ would have major implications for formation processes and favour models (e.g. Davies, Miller and Belovary 2011) with larger seed black holes ($M_{BH}\sim 10^5 M_{\odot}$), such as might form from the collapse of a  nuclear star cluster. 

Another reason why spectroscopy of $z>6$ QSOs is important is that these objects lie in the epoch of reionization, where (increasingly to higher redshifts) neutral hydrogen (HI) in the intergalactic medium absorbs photons emitted by the QSO blueward of Lyman-$\alpha$, especially at the redshift where they are shifted through this wavelength. Every QSO at these redshifts can be a useful probe of the HI content both near the source (e.g. Bolton et al. 2011) and along the line-of-sight.

 The UV and optical emission at $\lambda_{rest}>1216\rm \AA$ is not affected by HI absorption, but at these redshifts it is shifted into the near-IR where ground-based spectroscopy is very compromised by the atmospheric emission, so there are great efficiency gains from space-based observation. 
 Our aim in this paper is to investigate the prospects of identifying very high redshift QSOs in a {\it space-based wide-area near-infrared slitless spectroscopic survey}. We consider as an example the {\it Euclid} wide survey. 
 
  The European Space Agency mission {\it Euclid}, now scheduled for launch in 2019, will contain a Visible Instrument (VIS) and a near-infrared spectrometer and imaging photometer (NISP) sensitive at $11000<\lambda<20000\rm \AA$ (Laureijs et al. 2011). The latter instrument will be used to perform a slitless spectroscopic survey of the greater part of the sky outside of the Galactic plane, at least $15000$ $\rm deg^2$, over a period of several years. The principal target is emission-line star-forming galaxies at redshifts $0.67<z<2.07$ where the $\rm H\alpha$ line will be detected. By measuring redshifts (to $\sigma(z)<0.005$) for tens of millions of these galaxies, the large scale structure will be mapped in detail over an enormous volume (see e.g. Wang et al. 2010).

The survey is also expected to find $>10^4$ type 2 AGN and $\sim 10^6$ type 1 AGN. The moderate spectral resolution of NISP, $R\simeq 250$, should be enough to resolve the 
narrow $\rm H\alpha$ and $\rm [NII]6584$ emission lines for a high proportion of the galaxies (their flux ratio is an important diagnostic of type 2 AGN and also metallicity sensitive), and to recognize the broad emission lines characteristic of type 1 AGN. The broad lines are `permitted' transitions including $\rm H\alpha$, MgII and Lyman-$\alpha$ (see Fig. 1). Hao et al. (2005) found the $\rm H\alpha$ FWHM of low-redshift AGN to be bimodally distributed, $<1200$ km $\rm s^{-1}$ for type 2 AGN and  1400--8000 km $\rm s^{-1}$ for type 1s. Other broad lines in type 1 AGN spectra are in the same range (e.g. Sulentic et al. 2004; Shang et al. 2007), and in general greater than the FWHM of $\simeq 1200$ km $\rm s^{-1}$ corresponding to the NISP instrumental spectral resolution. 
{\it Euclid} NISP could be especially effective for identifying  QSOs at $z>8.06$, where the strong Lyman-$\alpha$ line would enter its wavelength range.

 We perform detailed simulations incorporating the most recent instrument parameters, a model of the QSO luminosity function (LF) and its evolution, and a spectral template based on observed high redshift QSOs, and follow with analysis of the simulated spectra. Section 2 describes the modelling of the QSO spectra and evolution, and Section 3 the set-up of the simulation. Section 4 describes the analysis of the spectra to obtain redshift and classifications and Section 5 presents our results, in terms of a QSO detection rate as a function of redshift and magnitude. In Section 6 we predict the numbers and some of the properties of the high-redshift QSOs which might be found in the NISP wide survey and Section 7 is summary and discussion. 

Magnitudes are given in the AB system where $m_{AB}=-48.60-2.5$ log $F_{\nu}$ (in ergs $\rm cm^{-2}s^{-1}Hz^{-1}$). We assume throughout a spatially flat cosmology with $H_0=70$ km $\rm 
s^{-1}Mpc^{-1}$, $\Omega_{M}=0.27$ and $\Omega_{\Lambda}=0.73$,  giving the age of the Universe as 13.88 Gyr.

\section{Model Representing the QSOs}
\subsection{Template Spectrum - Broad Lines}

To represent the intrinsic spectrum of all QSOs in the simulation we use the composite spectrum of Vanden Berk et al. (2001), which was created from over 2200 QSO spectra from the SDSS. The input spectra cover the redshift range $0.044 < z < 4.789$ and the full luminosity range  of the QSOs meeting the selection criteria. The composite covers 800--8555 $\rm \AA$ with a signal/noise peaking at $>300$ but much lower nearer either wavelength extreme. 

Table 1 lists the 7 most prominent broad ($\rm FWHM\simeq 2000$--6000 km $\rm s^{-1}$) lines in this spectrum, critical for identifying the spectrum as a QSO and determining the redshift. There are also narrow emission lines which have lower fluxes than these, the strongest being $\rm [OIII]5007$ ($\lambda_{vac}=5008.24\rm \AA$) with relative flux 2.49; there are many fainter lines, see Table 2 of Vanden Berk et al. (2001) for a full list. Fig. 1 shows the spectrum in the $F_{\lambda}$ form; the best-fit spectral index is $\alpha_{\nu}=-0.46$ ($\alpha_{\lambda}=-1.54$) over most of the range, becoming redder at $\lambda>6000\rm \AA$ due to host galaxy contributions.

At wavelengths shortward of Lyman-$\alpha$ at $1216\rm \AA$, and especially of Lyman-$\beta$ at $1030\rm \AA$, the Vanden Berk composite shows a drop in flux due to absorption from neutral hydrogen (HI) in the intergalactic medium (IGM). Fan et al. (2004) and Jiang et al. (2007) compare spectra of high luminosity ($M_B\simeq -27$) QSOs at $z\sim 6$ directly with this template and find good agreement, except the observed spectra fall off even more steeply in the blue wing of Lyman-$\alpha$ (making the line slightly narrower), because IGM absorption is stronger at $z\sim 6$ than at $z<4$. The spectral break at $\lambda_{rest}\simeq 1200\rm \AA$ becomes almost `total' e.g. Willott et al. (2010b) give a colour $i-z>3.07$ for a QSO at $z=6.44$.

Hence for our simulation it is preferable and more realistic to begin with an $unabsorbed$ QSO spectrum blueward of Lyman-$\alpha$, and then to subject this to absorption, modelled as a function of redshift.  At $\lambda_{rest}<1187\rm \AA$ we replace the Vanden Berk template by a power-law $f_{\nu}\propto\nu^{-1.76}$ ($f_{\lambda}\propto\lambda^{-0.24}$), as fitted by Telfer at al. (2002) to a mixed sample of low-$z$ QSOs (selected to be unaffected by absorption) at $500<\lambda<1200$. Additionally, the Lyman-$\beta$ line  (by far the strongest line in this range) is extracted from the Vanden Berk composite and re-inserted on top of the power-law. The resulting template, shown in Fig. 1, agrees well with the QSO composite in Fig. 4 of Telfer et al. (2002).
The template is extended redward (to the $K$ band) with the low-resolution mean QSO  spectrum (the all-QSO average) from Richards  et al. (2006), converted to $f_{\lambda}$ and spliced to the Vanden Berk spectrum at $\lambda\geq 6900\rm \AA$. 

\begin{figure} 
 \includegraphics[width=0.7\hsize,angle=-90]{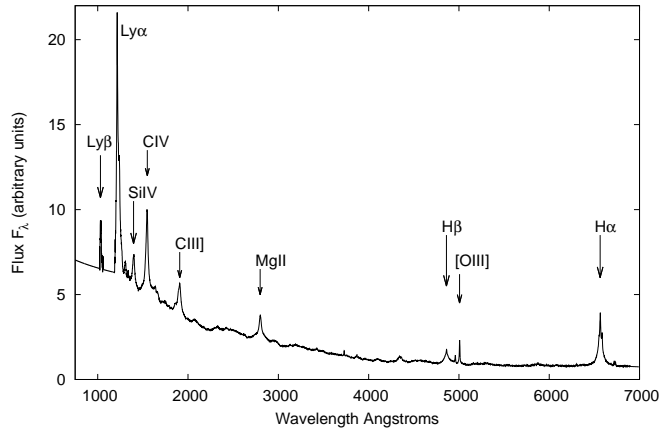}
\caption{Composite spectrum for type 1 QSOs in the SDSS (from Vanden Berk et al. 2001: replaced at $\lambda<1187\rm \AA$ by the power-law from Telfer et al. 2002), plotted on a linear scale, as flux per Angstrom ($F_{\lambda}$ in arbitrary units). Prominent broad lines are labelled, and also the strongest narrow line $\rm [OIII]5007$.} 
 \end{figure}

\begin{table}
\begin{tabular}{lccc}
\hline
Line & Wavelength ($\lambda_{vac}$) $\rm \AA$ & Relative flux & Width $\sigma_{\lambda}$ $\rm \AA$\\
\hline
Lyman-$\alpha$ & 1215.67 & 100.0 & 19.46\\
Si IV & 1396.76 & 8.9 & 12.50 \\
C IV & 1549.06 & 25.3 & 14.33 \\
C III] & 1908.73 & 15.9 & 23.58 \\
Mg II & 2798.75 & 32.3 & 34.95 \\
H$\beta$ & 4862.68 & 8.6 & 40.44 \\
H$\alpha$ & 6564.61 & 30.8 & 47.39 \\
\hline
\end{tabular}
\caption{Principal broad lines in the SDSS QSO template, with wavelengths in the vacuum}
\end{table}

For this simulation, we have assumed this averaged template (in which the Lyman-$\alpha$ rest-frame equivalent width $\rm EW\simeq 75\AA$) can approximately represent high-luminosity QSOs, even at $z>8$, provided we apply the appropriate IGM absorptions. This is an approximation as real QSO spectra are diverse and there are known examples of high-$z$ QSOs with very weak or absent Lyman-$\alpha$ lines (e.g. Fan et al. 1999, 2006). Also, some $z\sim 6$ QSOs have Lyman-$\alpha$ lines which are strong but much less broad, including the luminous J000552.34-000655.8 (Jiang et al. 2008; originally Fan et al. 2004) as well as less luminous ($M_B\simeq -25.5$) objects (Jiang et al. 2009). On the other hand, the lower-luminosity QSO studied by Kurk et al. (2009) at $z=6.08$ has a strong and broad Lyman-$\alpha$ (4350 km $\rm s^{-1}$; $\rm EW=87\AA$) even if its other lines are relatively narrow, and the highest redshift known QSO (Mortlock et al. 2011) has strong Lyman-$\alpha$.

Diamond-Stanic et al. (2009) investigate the  Lyman-$\alpha$ emission lines of high-redshift SDSS QSOs and find a mean EW of $67.6\rm \AA$ with a $1\sigma$ range 40--$102\rm \AA$. This 1-magnitude-wide range will introduce a corresponding scatter in the flux limits to which real QSOs can be identified but (as this is moderate and the distribution is centred at essentially the same EW as the adopted template) the total numbers of detectable QSOs should remain very close to our simulation prediction. But they also identified a population (20 objects forming $6\%$ of the QSOs at $z>4.2$) with extremely weak or absent Lyman-$\alpha$ emission ($\rm EW<15\AA$). These form an excess/tail above the EW distribution fitted to stronger-lined QSOs, impying that a few percent down-correction for a missed lineless fraction may still be needed for the predicted QSO counts (Section 6.1).

\subsection{Absorption from Intervening Neutral Hydrogen}
Emission from $\lambda<1216\rm \AA$ in the QSO rest-frame will be redshifted through Lyman-$\alpha$ along the line-of-sight to the observer and will be subject to absorption if neutral hydrogen is present. Shorter wavelengths will also be subject to absorption by Lyman-$\beta$ etc. The density of HI strongly increases with redshift. In our simulation, the intrinsic QSO template is subjected to absorption of optical depth $\tau(\lambda,z)$, following the model of Madau (1995).

Photons blueward of Lyman-$\alpha$, emitted from the QSO at redshift $z_{q}$ will be observed with wavelength $\lambda_{obs}$ where $\lambda_{obs}<\lambda_{\alpha}(1+z_{qso})$ and $\lambda_{\alpha}=1216\rm \AA$. They suffer Lyman-$\alpha$ absorption along the line of sight at a redshift $z_{abs}=\lambda_{obs}/\lambda_{\alpha}-1$ (always $z_{abs}<z_{qso}$), with an optical depth (Madau 1995) of approximately $\tau_{\alpha}=0.0036(1+z_{abs})^{3.46}$.

There are further additive terms in $\tau$ (of decreasing size) for wavelengths emitted shortward of Lyman-$\beta$ ($\lambda_{\beta}=1026\rm \AA$), Lyman-$\gamma$ and further lines in the series. Our model  includes the Lyman-$\alpha$, $\beta$, $\gamma$ and  $\delta$ absorption terms as given by Madau (1995), ignoring the higher lines, but including the absorption of $\lambda_{em}<912\rm\AA$ (beyond the Lyman limit) photons with the optical depth given by the more complex analytic expression in the footnote on page 21 of Madau (1995).
Note the simulation is insensitive to the details of the model used for HI absorption, as its effects will become visible to NISP only at $z>8.06$, where the model gives a very strong break with almost zero flux at $\lambda_{rest}<1200\rm \AA$; something already seen for QSOs at $z>6$ (e.g. Willott et al. 2011b).

Here the Madau (1995) absorptions are only applied in full up a distance $\Delta(z)=z_{qso}-z_{abs}=0.125$ away from the QSO. This is to represent the ionized HII `bubble' (Str{\"o}mgren sphere) of several Mpc radius that a QSO forms around itself by its own ionizing-photon emission.The effect of this is to allow the light in the Lyman-$\alpha$ line itself, or most of it, to reach the observer, even if there is very heavy absorption of wavelengths immediately blueward of the lines (e.g. Becker et al. 2001).
Carilli et al. (2010) study the ionization zones for a sample of $5.7<z<6.4$ QSOs, plotting transmission vs. radius profiles (which appear approximately Gaussian), and define a zone radius as that where photon transmission falls to $10\%$.

In our model, at $\Delta(z)<0.125$ from the QSO, the Madau model optical depth from each line is reduced to  $\tau=[(z_{qso}-z_{abs})/0.125]^2$, which gives a Gaussian-profile decline in transmission versus radius (or equivalently, in wavelength shortward from the emission line). Transmission is  $100\%$ at $\lambda_{rest}=1216\rm \AA$ but, for a QSO at $z=6$, falls to $10\%$ at a proper radius 7.3 Mpc (in agreement with Carilli et al. 2010). At this radius, emission from the QSO at $\lambda_{rest}=1196\rm\AA$ will be redshifted through Lyman-$\alpha$, so this is approximately the rest-frame wavelength the observed spectrum would show a cut-off (break).
Fig. 2a shows the modelled transmission as a function of rest-frame wavelength for 
QSOs at different redshifts.

 Carilli et al. (2010) and Bolton et al. (2011) find the ionized zone radii tend to decrease with increasing redshift, a sign that the (late end of the) IGM reionization epoch has already been reached at $z\simeq 6$. Our model's  ionized zone radius decreases with redshift as, firstly, the proper radius corresponding to $\Delta(z)=0.125$ decreases with redshift, and in addition, the optical depth at
$\Delta(z)=0.125$ increases rapidly, as $(1+z)^{3.46}$, causing the radius where transmission is $10\%$ to decrease more steeply (Fig. 2b). The ionized zone radius for the $z=7.085$ QSO of Mortlock et al. (2011) is only 1.9 Mpc (so if typical might suggest even faster evolution).   In our model this radius is still 1--3 Mpc in the $8<z<10$ range of most interest for this simulation, consistent with evidence that $\geq 1$ Mpc scale ionized zones can already exist for galaxies at $z\sim 8.5$  (Lehnert et al. 2010).

\begin{figure} 
 \includegraphics[width=0.7\hsize,angle=-90]{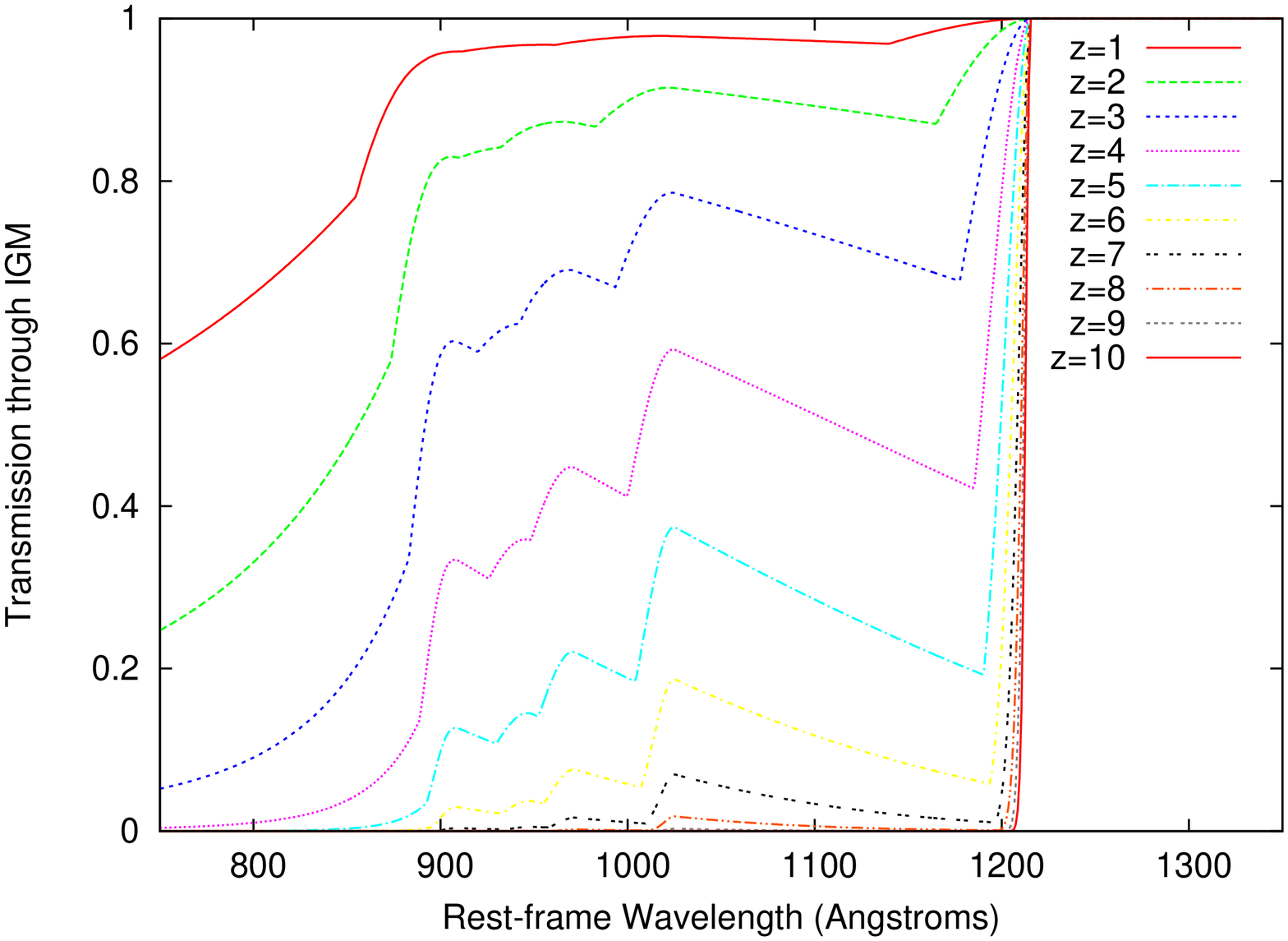}
\includegraphics[width=0.7\hsize,angle=-90]{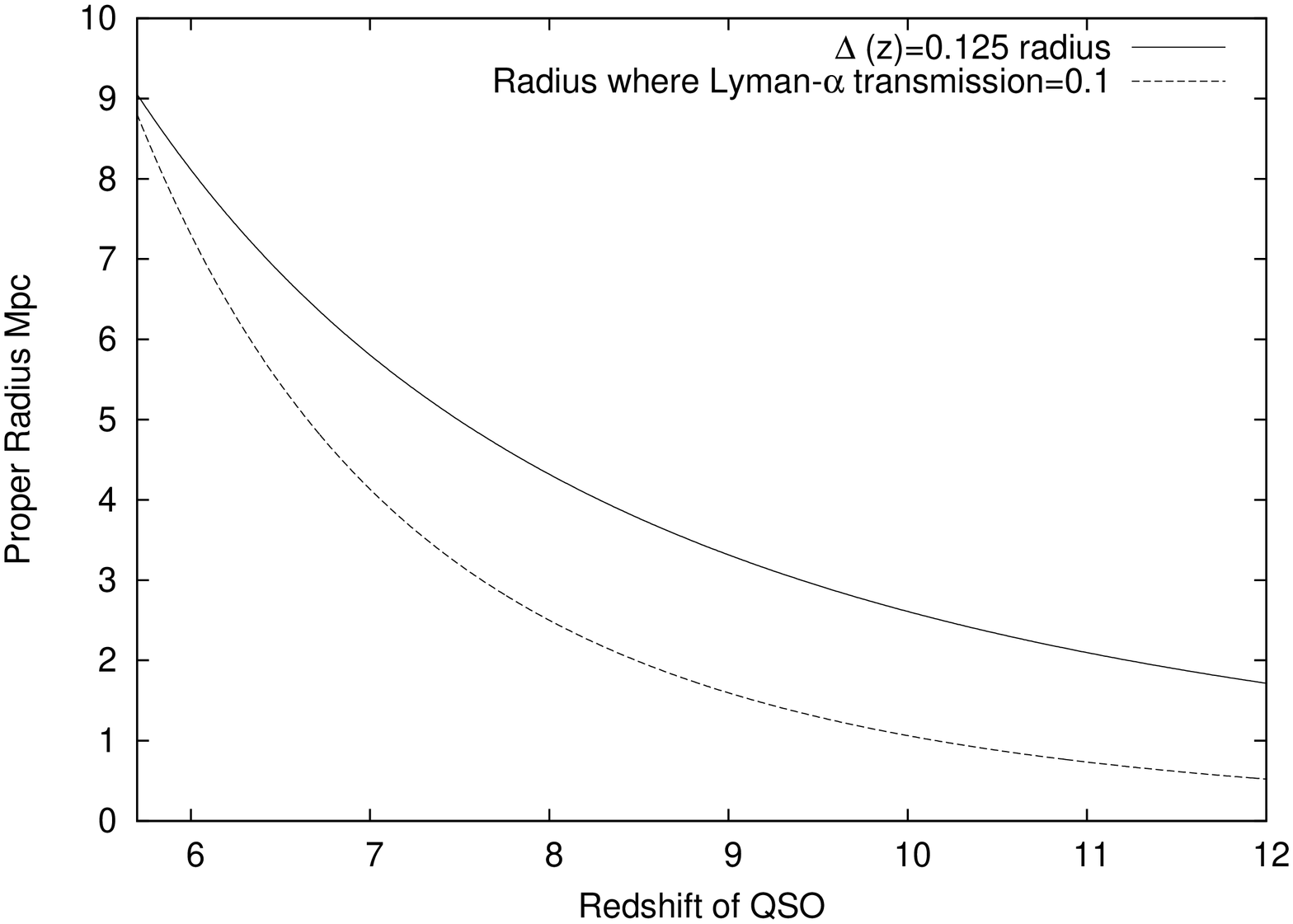}
\caption{(a) Model (from Madau 1995) for IGM line-of-sight transmission as a function of wavelength in the QSO rest-frame, for different QSO redshifts, showing increasing absorption shortward of Lyman-$\alpha$; (b) The proper radius from the QSO corresponding to $\Delta(z)=0.125$ and where (in our model) the transmission of light being redshifted through the Lyman-$\alpha$ wavelength falls to 0.1 (10\%), as a function of QSO redshift.} 
 \end{figure}

Fig. 3 shows the QSO template spectrum in the Lyman-$\beta$ to Lyman-$\alpha$ range, at a range of redshifts, after applying the modelled IGM absorption. There will be almost zero flux observed at wavelengths shortward of Lyman-$\alpha$ for all $z>7$ sources. We expect the absorption to `bite' increasingly into the blue wing of the line as redshift increases, making it narrower and more asymmetric in profile, but as the absorption (in this model) only ever affects the blue side, Lyman-$\alpha$ should still remain strong and detectable. 
\begin{figure} 
 \includegraphics[width=0.7\hsize,angle=-90]{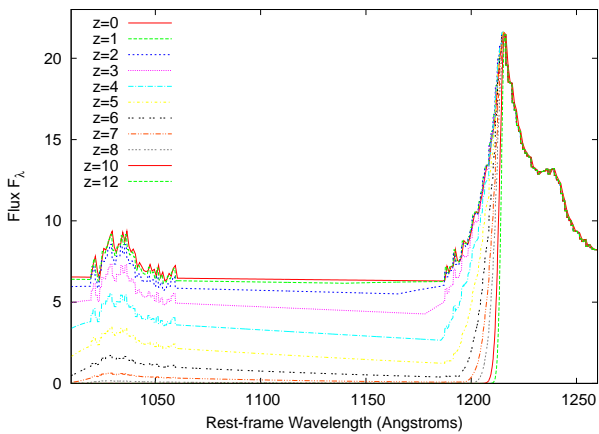}
\caption{The QSO template after IGM absorption, plotted in rest-frame $\lambda$ and showing the  Lyman-$\beta$--$\alpha$ range only, illustrating the marked effect of increasing absorption on the Lyman-$\alpha$ line profile and spectral break.} 
 \end{figure}
\subsection{Luminosity Function and Evolution}
The LF of AGN is usually represented by a double power-law form (Boyle et al. 2000), expressed in absolute magnitudes as:

$$\Phi(M,z)={{\Phi^*_M}\over{10^{0.4\alpha(M-M^*)}+10^{0.4\beta(M-M^*)}}}$$

The bright- and faint-end slopes are respectively parametrized as $\alpha$ and $\beta$, with $\alpha$ being steeper. The AGN LF undergoes a strong evolution with redshift, which has often been represented by a pure luminosity evolution model (PLE), e.g. the form $M_g=-22.18-1.44z+0.315z^2$ from Croom et al. (2009), in which the $L^*$ luminosity rises to a peak at $z\simeq 2.3$ and decreases again beyond this. 
However, the PLE model is not the best-fitting model, as there is evidence that the evolution is luminosity dependent, with the flux from lower luminosity AGN peaking at lower redshifts. Two ways of representing this are (i) a model in which both $L^*$ and the bright and/or faint-end slopes of the LF evolve with redshift (e.g. Croom et al. 2009), and (ii) luminosity-dependent density evolution (LDDE). 
In this paper we use the LDDE model as given by Bongiorno et al. (2007), which was fitted to type 1 AGN from the VIMOS VLT Deep Survey (VVDS) and SDSS over the range $1<z<4$.

                                                                                                                                                                                                                                                                                                                                                                                                                                                                                                                                                                                                                                                                                                                                                                                                                                                                                                                                                                                                                                                                                                                                                                    However, at higher redshifts ($z\sim 4$) both optical and X-ray data indicate more rapid negative $\Phi^*$ evolution than given by this model.
Schmidt et al. (1995) and Brusa et al. (2009) found observations of QSOs at $z=3.0$--4.5 to be consistent with an exponential decrease in comoving number density,
 $\Phi\propto e^{-1.0(z-2.7)}$ or $10^{-0.43(z-2.7)}$. Civano et al. (2011) found this same form to fit {\it Chandra}-COSMOS counts to even higher redshifts of $z\sim 5.3$. The LF analysis of Willott et al. (2010a) assumed slightly faster evolution of  $10^{-0.47z}$ at $z\sim 6$ and Fontanot et al. (2007a) fitted the QSO counts at $3.5<z<5.2$ with evolution  $e^{-1.26(\pm0.18)z}=10^{-0.55(\pm 0.08)z}$; the differences between these estimates are probably only at the level of the statistical uncertainties. 

We therefore use the Bongiorno et al. (2007, hereafter B07) LDDE model for $0\leq z\leq 2.7$ only, and at all $z>2.7$ assume density evolution with a rate $\Phi(M_B,z)=\Phi(M_B,2.7)10^{-0.47(z-2.7)}$, calling this model $k_{ev}=-0.47$. Fig. 4 shows the resulting $B$-band LF at redshifts $z=0$ to $z=12$. 

For comparison, also plotted is the LF fitted by Willott et al. (2010a) to a sample of $z\simeq 6$ QSOs, which is $\Phi^*=1.14\times 10^{-8}$, $\alpha=-1.81$, $\beta=-0.5$ and $M_B=-25.81$ (assuming $B=m_{1450}-0.68$ as calculated from the template spectrum). At $z=6$ the Willott et al. (2010a) LF is a little below our model, by $40\%$ at $M_B=-28$, and it is also $\simeq 40\%$ lower in normalization than the Jiang et al. (2009) LF, which was fitted to (high luminosity) $z\simeq 6$ QSOs in the SDSS. The sample of Willott et al. (2010a) includes these same QSOs (so is not entirely an independent estimate) but it is more weighted to lower luminosity ($M_B\simeq -25.4$) QSOs in the CFHQS. At this time more $z>6$ QSOs are still needed to better constrain their LF at these redshifts; in Section 6.1 we will consider both normalizations.
\begin{figure} 
 \includegraphics[width=0.7\hsize,angle=-90]{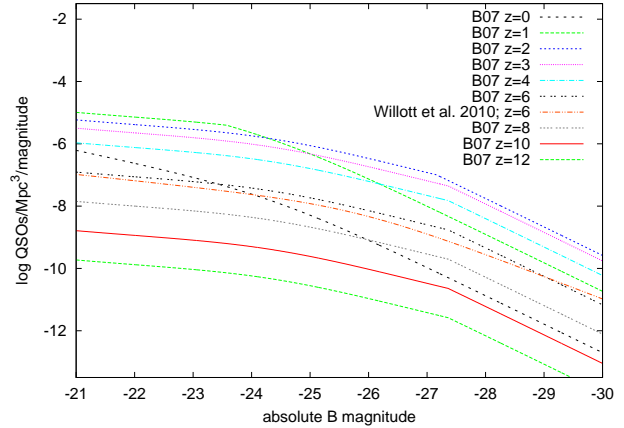}
\caption{Our model luminosity function (B07 LDDE plus $k_{ev}=-0.47$ density decrease at $z>2.7$) in the $B$-band for Type 1 QSOs at a series of redshifts, showing a rise from $z=0$ to $z=2$--3 followed by a steady decine to $z=12$. Also shown is the Willott et al. (2010a) best-fit LF for QSOs at $z\simeq 6$ (assuming $B=m_{1450}-0.68$).} 
 \end{figure}

Fig. 5 shows the corrections $J_{observed}-B_{rest}$, $H_{observed}-B_{rest}$ and $K_{observed}-B_{rest}$ calculated for our model's QSO template spectrum (including the redshift-dependent absorption shortward of Lyman-$\rm alpha$). Also the $B$-band k-correction, approximately $-1.35~{\rm log}(1+z)$ to $z\simeq 4$, beyond which absorption causes the source to drop out of $B$-band observations, while remaining visible in the near-IR bands to $z>10$. For the simulation we consider primarily the apparent magnitudes in the $H$-band, 1.40--$1.80\mu \rm m$, the passband most similar to the NISP spectral range. The colours between the NIR passbands remain small, at least to $z\sim 10$; e.g at $z=8.2$, $H=21.0$ corresponds to $K=20.90$ and $J=21.28$.

\begin{figure} 
 \includegraphics[width=0.7\hsize,angle=-90]{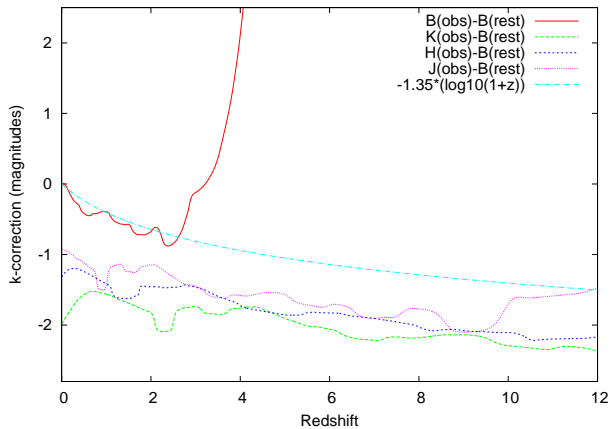}
\caption{Blue-band k-correction from $z=0$ to $z=12$, and the corrections $J_{observed}-B_{rest}$, $H_{observed}-B_{rest}$ and $K_{observed}-B_{rest}$, computed for the QSO composite spectrum (including IGM absorption).} 
 \end{figure}

Fig. 6 shows the model's redshift distribution, N(z), for type 1 QSOs at a series of $H$ magnitude limits. At all these limits the $N(z)$ peaks at $z\sim 1$ and very high  redshift QSOs remain very rare, e.g. to $H=22$ the fraction at $z>8.06$ is only $0.0016\%$ (although for a  15000 deg$^2$  area this still amounts to some 86 objects). 

\begin{figure}
 \includegraphics[width=0.7\hsize,angle=-90]{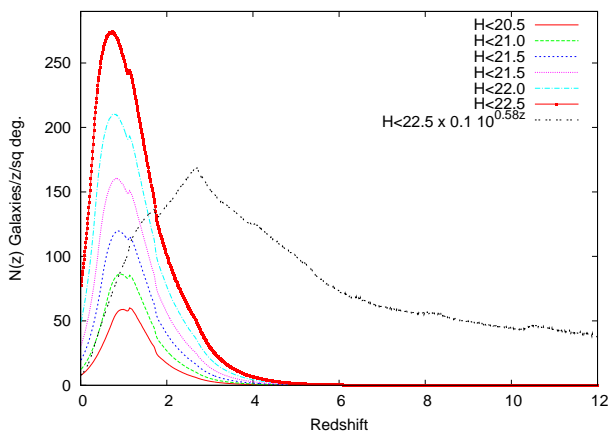}
\caption{Our model redshift distribution $N(z)$ for QSOs at limits $H=19.5$ to 22.5.
Also plotted (multiplied by 0.1 to fit on the same graph) is the $H\leq 22.5$ $N(z)$ with a scaling up by $10^{0.58z}$, to greatly increase the proportion of high redshift AGN; this is the $N(z)$ used for the simulation input catalog (Section 3).}
\end{figure}
 
We also compare with predictions for $5<z<12$ QSOs tabulated by Fontanot, Somerville and Jester (2007b, hereafter F07b) for a deeper $H\leq 23$ limit. Fig. 7 shows our model QSO counts in
 $\Delta(z)=0.5$ bins compared with the F07b non-evolving, $k_{ev}=-0.54$ exponential evolution (from Fontanot et al. 2007a) and Eddington-accretion models, all computed with the `13b' LF. They are very similar to  the F07b exponential model. The non-evolving model gives a very much higher count (and is already excluded by data, so is shown for comparison only), while the Eddington accretion model predicts a very steep decline with QSOs only visible to $z\sim 7.5$, even at this faint limit. This is an extreme Eddington accretion model in which the QSOs have constantly accreted at the Eddington limit since formation up to $z=6$. i.e. with a maximal duty cycle of 1.
 
Similarly in the `reference model' of Shankar et al. (2010) the AGN duty cycle is very close to unity at $z\geq 6$ and their numbers (at a given luminosity) decline almost twice as steeply as in our model. 
With a lower duty cycle the growth of QSOs is more protracted so that more are detectable at $z>8$ and some might be visible at $z>10$ (e.g. Fig.10 of Rhook and Haehnelt 2008).

\begin{figure}
 \includegraphics[width=0.7\hsize,angle=-90]{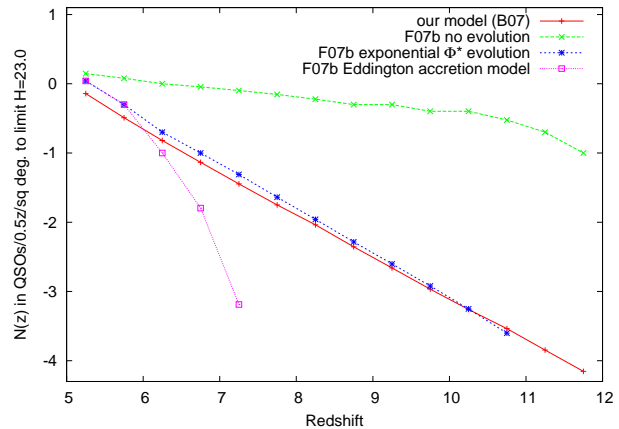}
\caption{Predicted counts of high-$z$ QSOs in $\Delta(z)=0.5$ intervals, shown on a log scale for a limit $H\leq 23.0$. Our model is compared with the non-evolving, exponential evolution ($k_{ev}=-0.54$) and Eddington accretion models of F07b (for the 13b LF).}
\end{figure}

\section{The Simulation}
\subsection{Set-up}

We generate simulations of NISP slitless spectroscopy images using {\sevensize AXESIM} (Rosati et al. 2009; Franzetti et al., in preparation).
The necessary inputs for the simulation include the QSO template spectra and a catalog of QSO magnitudes and redshifts. We use a set of 61 templates to represent the QSOs; they are identical at $\lambda>1216\rm \AA$ but shortward of Lyman-$\alpha$ include differing levels of IGM absorption (modelled as described above) corresponding to $\Delta(z)=0.2$ intervals from $z=0$ to $z=12$. 

As we aim to estimate detection efficiency over all redshifts, the $N(z)$ used to set up the simulation catalog is modified from the (B07) model by a scaling factor $10^{0.58z}$, giving a greatly increased proportion of high redshift objects - almost a flat $N(z)$ (Fig. 7) - while retaining the `correct' apparent magnitudes (as given by the model) for each QSO.  
As type 1 QSOs in the wavelength range of interest are typically 85--90\% dominated by the central AGN (e.g. Hutchings et al. 2002, Goto et al. 2009), all the QSOs  are represented as point sources (their size on the simulation image is just the instrumental point-spread function).

{\sevensize AXESIM} generates a simulation image containing the spectra of the input catalog objects, exactly as would be seen by NISP using the most recent instrumental parameters (Laureijs et al. 2011). We simulate a single square area $0.5\times 0.5$ $\rm deg^2$, and on this are placed a total 2187 QSO spectra, with redshifts up to $z=12$ and to the faint limit $H=22.5$ (preliminary tests found no QSOs fainter than this would be detected).

 For the simulation to be realistic it must at least attempt to replicate the effect of spectral contamination by neighbouring spectra, a serious problem for deep slitless spectroscopy, which will reduce the numbers of spectra correctly identified (i.e. survey efficiency). Contamination is introduced in part by the artificially very high surface density of high-redshift QSO spectra (2187; without the $10^{0.58z}$ term there would only be 120) and partly by adding spectra of non-AGN faint galaxies. The galaxy spectra are copied directly from a $0.5\times 0.5$ $\rm deg^2$ area of our COSMOS-IRAC  (Spitzer Infrared Array Camera) based simulation of galaxies in NISP (Franzetti et al., in preparation), but because there is already a greatly enhanced surface density of AGN spectra, only every other galaxy spectrum from this model is added. This still amounts to 13023 faint galaxies on the frame, although only 5814 of these are brighter than the simulation limit ($K\simeq 22.4$) and only 501 have $\rm H\alpha$ emission lines above the nominal survey detection limits
($0.67<z<2.05$ and $F(\rm H\alpha)>3\times 10^{-16}$ ergs $\rm cm^{-2}s^{-1}$). This representation of spectral contamination is inevitably approximate, as for one thing the emission-line properties of the simulated galaxies are significantly model-dependent, and because the simulated broad-line QSOs and the galaxies differ greatly in spectra and redshift ranges and so cannot produce quite equivalent effects. But it should be good enough for a reasonable estimate contamination's adverse effects on QSO redshift measurement and for of us to test a method of reducing this (Section 4.3).

 The total exposure time for the simulation image is 1600 sec in total, which,  following the currently planned wide survey configuration is split into 4 exposures -- the 4 combinations of two (blue-half and red-half) filters and two slit roll angles ($0^\circ$ and $90^\circ$). 
\subsection{Extracting the Spectra}
After generating the simulation image, {\sevensize AXESIM} then extracts the spectra for each target object, at both roll angles, exactly as would be done for real NISP data (K{\"u}mmel et al. 2009). The spectra each have 919 pixels covering $11000<\lambda<20000\rm \AA$ with pixelsize $9.8\rm \AA$. We will here be considering only the QSO spectra. Because of the limited field size, many spectra are partially off the field edges. In total, because of geometric considerations, only 1468 of the QSO spectra could be extracted successfully (including 37 for which we obtained only one of the two roll angles). However, losses of this type will be far less significant in the real wide survey, in which {\it Euclid} will scan strips or patches with the fields-of-view mosaicked to cover much larger contiguous areas without gaps (Laureijs et al. 2011). Hence for the purposes of this paper we disregard the spectra that could not be extracted (because they are partially outside the field of view), and successful detection rates will be calculated as fractions of the numbers of extracted spectra. 

For each observed object  1D spectra from the zero and $90^{\circ}$ roll angles were combined making use of a decontamination tool which is part of {\sevensize RESS}
(Redshift Extraction from Slitless Spectroscopy) an automatic software
running within IRAF environment, which has been devised (Rossetti E. 2011, in
preparation) to reduce and analyze contaminated slitless spectra.
{\sevensize RESS} $z$-measure is currently implemented only for low $z$ galaxies ($0.7 < z
< 2.0$) and its extension for high $z$ objects is foreseen.  The {\sevensize RESS} decontamination
tool is optimized to detect and exclude from the final combination the
contaminated portions in each spectrum, when present. However, this is only sometimes effective where, as in this case, there are only 2 roll angles, and will be much more efficient for the deep fields where a greater number of roll  angles will be utilized.

Fig. 8 shows 6 examples of spectra extracted from the simulation (roll angles summed); these are brighter ($H<19$) examples without significant contamination, in order to most clearly illustrate the spectral features visible at different redshifts. 
\begin{figure} 
 \includegraphics[width=0.7\hsize,angle=-90]{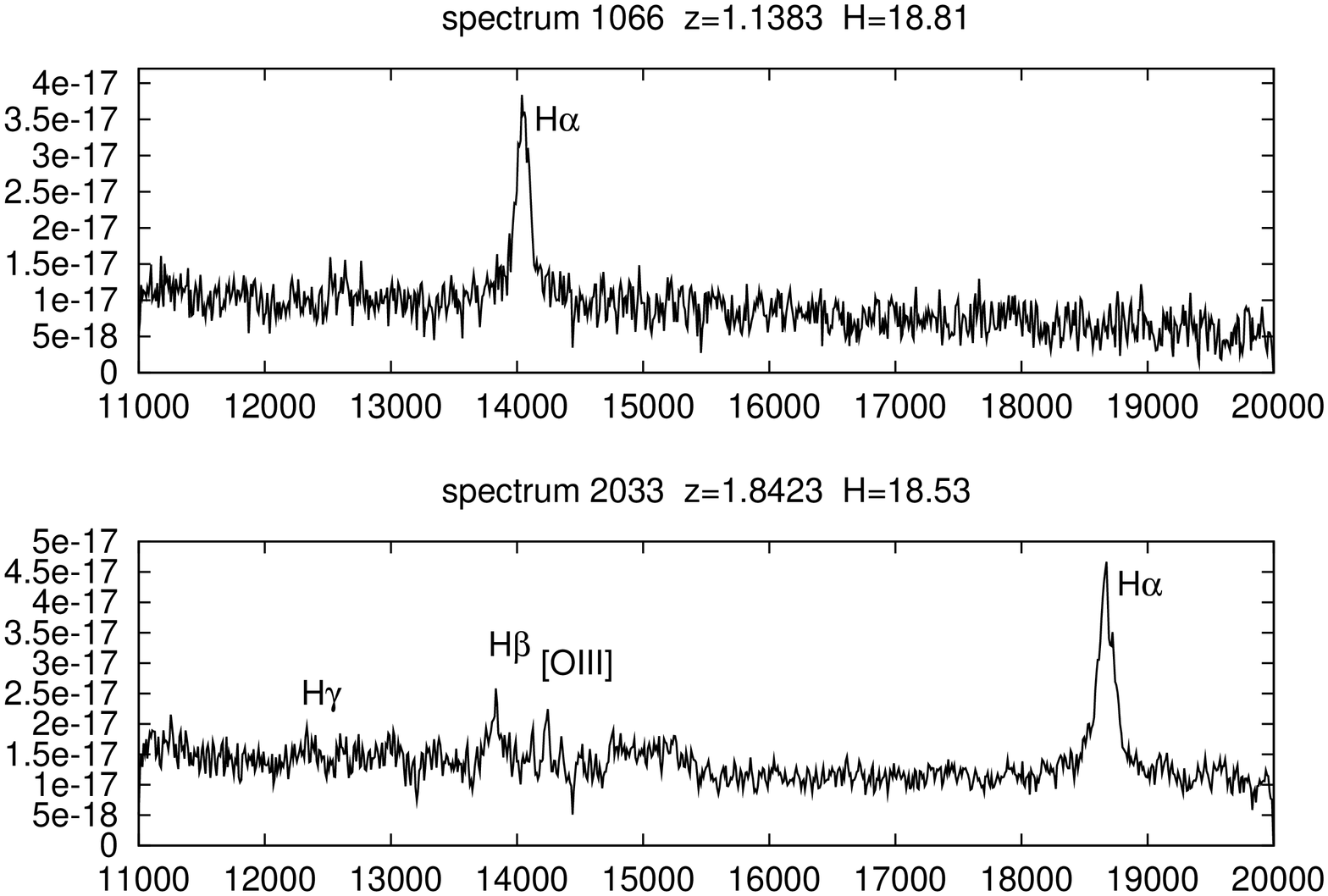}
 \includegraphics[width=0.7\hsize,angle=-90]{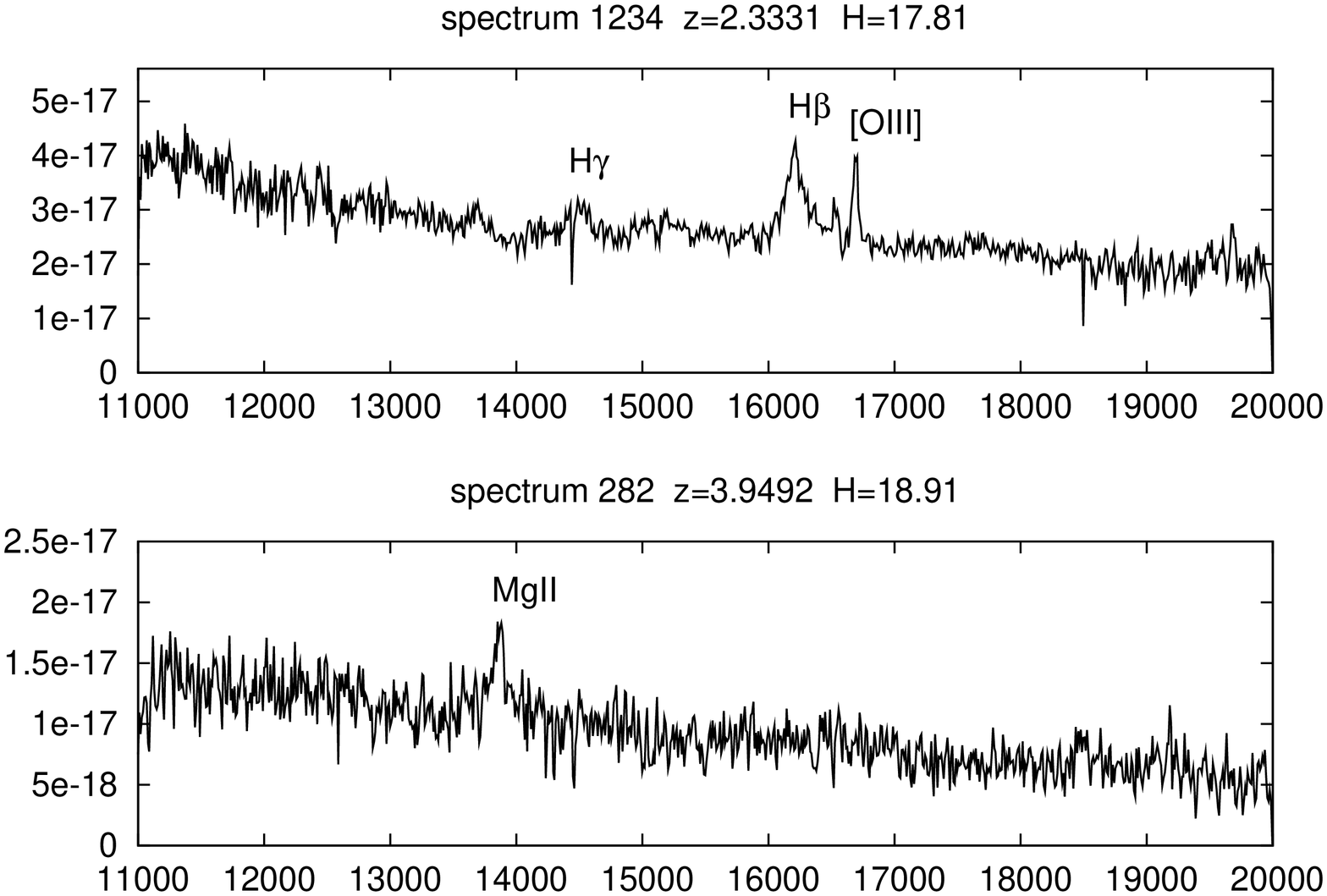}
\includegraphics[width=0.7\hsize,angle=-90]{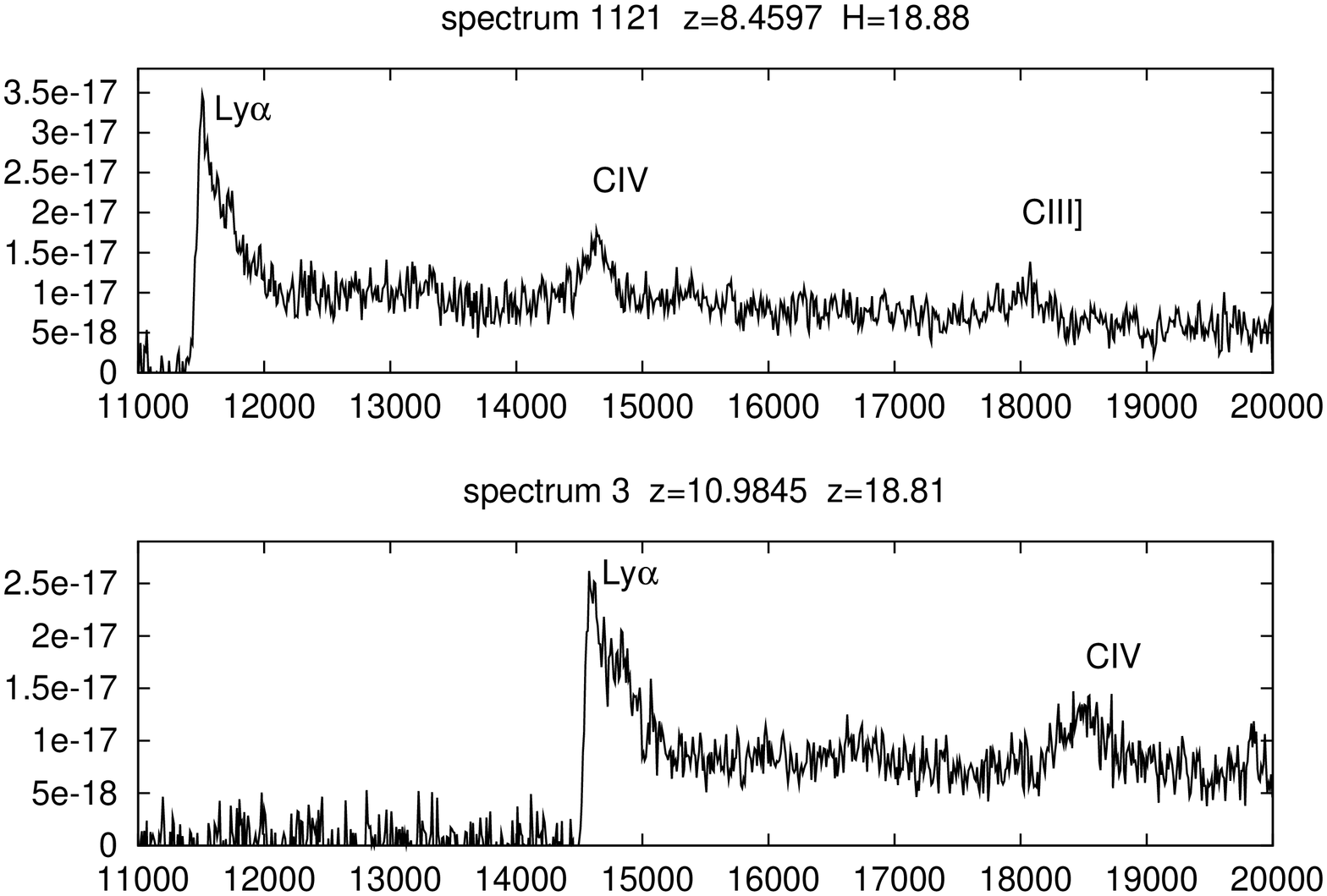}
\caption{Simulated spectra plotted as flux $F_{\lambda}$, units ergs $\rm cm^{-2}s^{-1}\AA^{-1}$, against $\lambda$ in Angstroms, as extracted from the simulation image (afer combination of the two roll angle spectra). These are six of the brighter ($H<19$) examples at different redshifts, with the strongest visible emission features labelled. The last two are at $z>8.06$ where Lyman-$\alpha$ enters the NISP range.}
\end{figure}

\section{Analysis of the Simulated Spectra}
\subsection{Method: Automated Redshift Finder plus Line Width Measurement}
For a survey of this size the spectra must (first) be analyzed with a fast and efficient automated redshift finder, and for our simulation we use the program {\sevensize EZ} (Garilli et al. 2010), which is supplied with a large set of template spectra, including the SDSS-composite QSO spectrum and a wide variety of emission-line and absorption-line galaxies.  {\sevensize EZ} classifies spectra and fits redshifts by a procedure which combines (i) correlating spectra with templates (ii) refining the correlation with a $\chi^2$ fit (which can often choose the best amongst several redshifts where the correlation peaks) and (iii) emission line finding and matching, only in cases where there are emission lines detected above a chosen threshold (by default $5\sigma$, which is retained here).

For our analysis we add five more templates, the Vanden Berk et al. (2001) composite modified shortward of Lyman-$\alpha$ (as described in Section 2.1 and plotted on Fig. 1), and the same but with the IGM absorption (as modelled in Section 2.2) appropriate to $z=2,4,6$ and 8, giving an increasingly sharp and deep spectra break and asymmetric Lyman-$\alpha$ line (Fig. 4). Adding these templates is important, they improve the performance and accuracy of redshift measurement at redshifts where the Lyman-$\alpha$ is visible.
 
{\sevensize EZ} was run in batch (automated) mode on the full set of 1468 QSO spectra extracted from the {\sevensize AXESIM} image, outputting for each a best-fit redshift and template (i.e. a spectral classification). To evaluate the results we count the numbers of  spectra for which the input and best-fit redshifts agree within a chosen tolerance and, in addition, the spectrum is correctly classified (i.e. the best-fit template is any QSO template, rather than e.g. a starburst galaxy).

However, there is a known `problem' with the current {\sevensize EZ} configuration, that if multiple $>5\sigma$ lines or any `strong' ($>10\sigma$) lines are found, the spectrum is only correlated with emission-line galaxy templates.  While this feature might be removed, this might lead to undesirable misidentification of some real emission-line galaxies. A better solution is to follow redshift-measurement by a further analysis to fit the width (FWHM) of the strongest emission line. Any spectrum found to have a bright line of width $\rm FWHM>2000$ km $\rm s^{-1}$ can be confidently reclassified as a type 1 AGN. It will also be useful to investigate here the limits to which line FWHM can accurately be measured with NISP data, as they can provide information on the AGN properties, including (at least in the case of $\rm H\alpha$, $\rm H\beta$ and MgII) the mass of the SMBH (e.g. Trakhtenbrot et al. 2011, Willott et al. 2010b)

The present version of {\sevensize EZ} does not include line-width measurement, and so this analysis is performed with {\sevensize IRAF} `specfit', fitting each spectrum with a simple two-component model -- a power-law continuum and a single Gaussian line (with FWHM from 500 to 10000 km$\rm s^{-1}$). 
The returned parameters (4) are power-law normalization and slope, Gaussian flux and FWHM, all with error bars. The line to be fitted in each spectrum is chosen as the brightest broad line visible (for $0.67<z<2.04$, $\rm H\alpha$; for $2.04<z<3.1$, $\rm H\beta$; for $3.1<z<6.128$, MgII; for $6.128<z<8.1$ CIV, and 
for $z>8.1$, Lyman-$\alpha$), and the Lyman-$\alpha$ fit was allowed to be skewed (in the range 0.8--1.3) and the other lines assumed to be symmetric.

 \subsection{Results: Redshifts and Classifications}
For spectra such as these with broad lines and low signal-to-noise ratio, redshift measurements are likely to be less precise than for galaxies with strong narrow lines, and we allow here a wide tolerance of $\Delta(z)=\pm 0.05$ for a redshift to be considered `correct'. The scatter will be examined below.

Examining the automated output for all 1468 analysed spectra, we can sort the results as follows:

(i) spectrum is classed as an QSO with a correct redshift (148 spectra),

(ii) correct redshift but classed as a non-AGN galaxy (20 spectra),

(iii) incorrect redshift, classed as AGN (171 spectra),

(iv) incorrect redshift, classed as a non-AGN galaxy (1129 spectra).

A plot of $\Delta(z)=z_{fit}-z_{input}$ against redshift (Fig. 9) shows that that for spectra in the Lyman-$\alpha$ window there is a relatively large scatter between these two redshifts. For spectra with correct ($\Delta(z)<\pm 0.05$) redshift measurements in this $z$ range (numbering 42) the scatter $\sigma(z_{fit}-z_{input})=0.0237$ and this is also skewed, with a systematic offset $\langle z_{fit}-z_{input}\rangle=0.015\pm0.004$. This skewness must be due to the very asymmetric form of the Lyman-$\alpha$ line and break. In the $\rm H\alpha$ window the (68) spectra identified as AGN with correct $z$ have a much smaller redshift scatter $\sigma(z_{fit}-z_{input})= 0.0026$ with no systematic offset (as $\rm H\alpha$ is narrower and symmetric).

\begin{figure}
 \includegraphics[width=0.7\hsize,angle=-90]{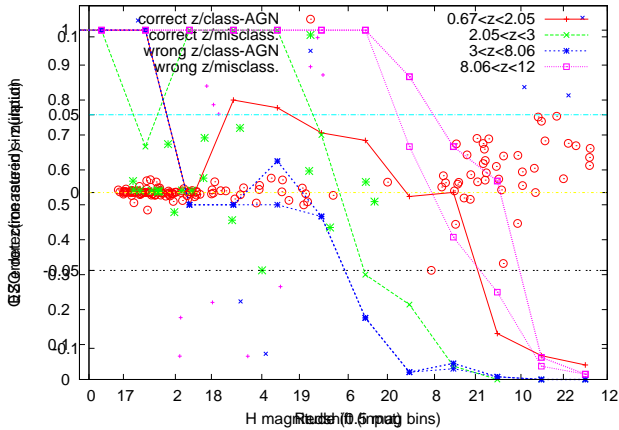}
\caption{The difference between input redshift $z_{in}$ for a spectrum and the redshift best-fitted by {\sevensize EZ}, as a function of $z_{in}$, showing the tolerance $\Delta(z)=\pm 0.05$. Note the redshift-dependence of the scatter.}
\end{figure}

Of the 20 spectra with correct redshifts, but incorrect classifications as non-AGN galaxies, none are at $z>8$, most are at $1<z<3$, and 7 are relatively bright with $H<20$ with an 8th only slightly fainter with $H=20.058$. In 2 of these 8 brighter spectra the misidentification was due to contamination distorting a region of the continuum (if these regions are excluded from the fit, the correct AGN classification is obtained). In the other 6 cases, {\sevensize EZ} identified multiple emission lines above $5\sigma$, which could be $\rm H\alpha$, $\rm H\beta$ and one or both of the $\rm [OIII]$ lines (5007, 4959). Many can be reclassified as QSOs using line width measurements.

As a correct redshift is required beforehand in order to set an initial $\lambda$ for the line to be fitted, we perform the `specfit' analysis only on the 168 spectra for which {\sevensize EZ} found a correct redshift.
 Fig. 10 shows examples of the specfit fits. Fig. 11 shows the measured FWHMs as a function of $H$ magnitude for $\rm H\alpha$, $\rm H\beta$, MgII and Lyman-$\alpha$ lines.

\begin{figure} 
 \includegraphics[width=0.7\hsize,angle=-90]{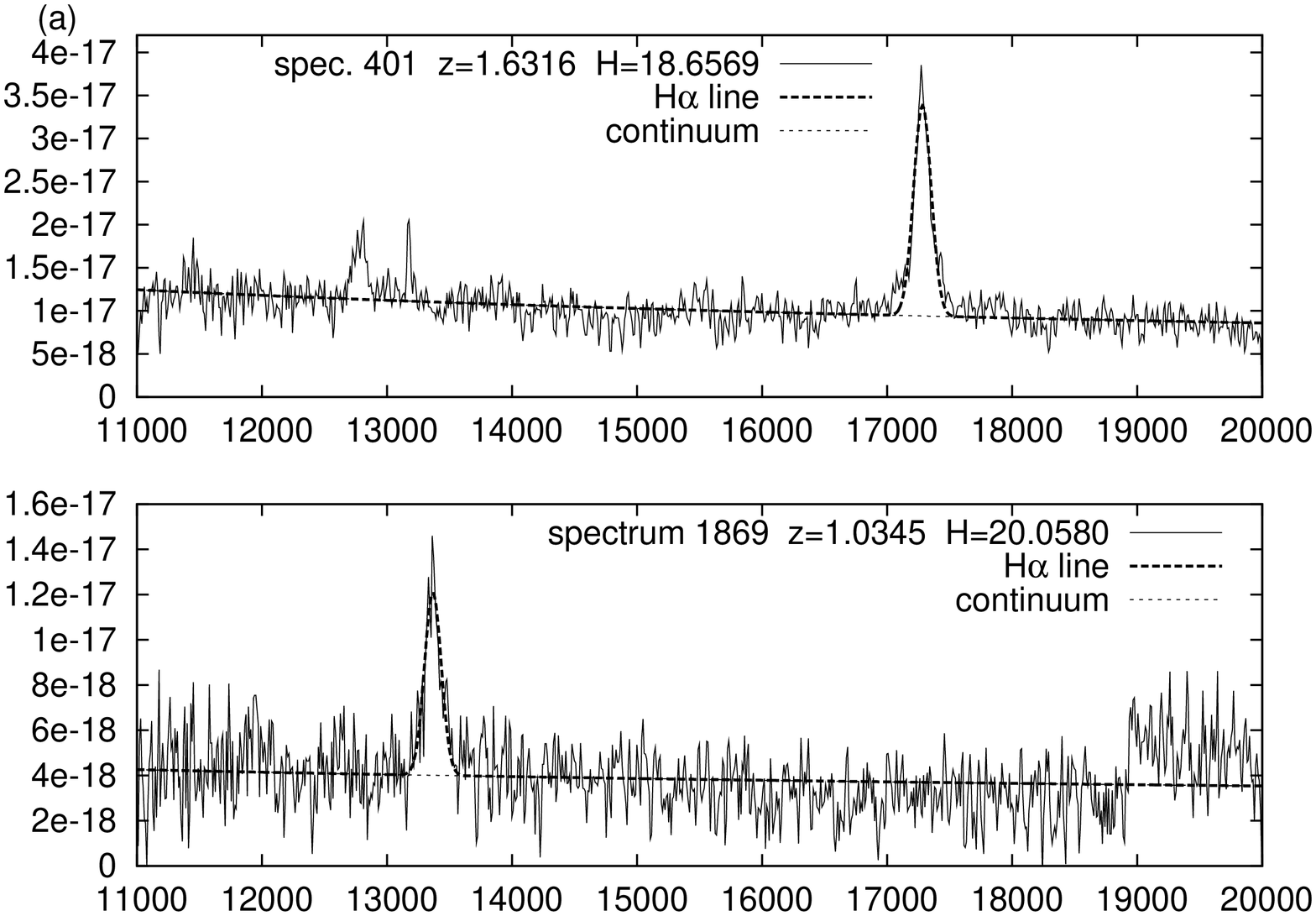}
  \includegraphics[width=0.7\hsize,angle=-90]{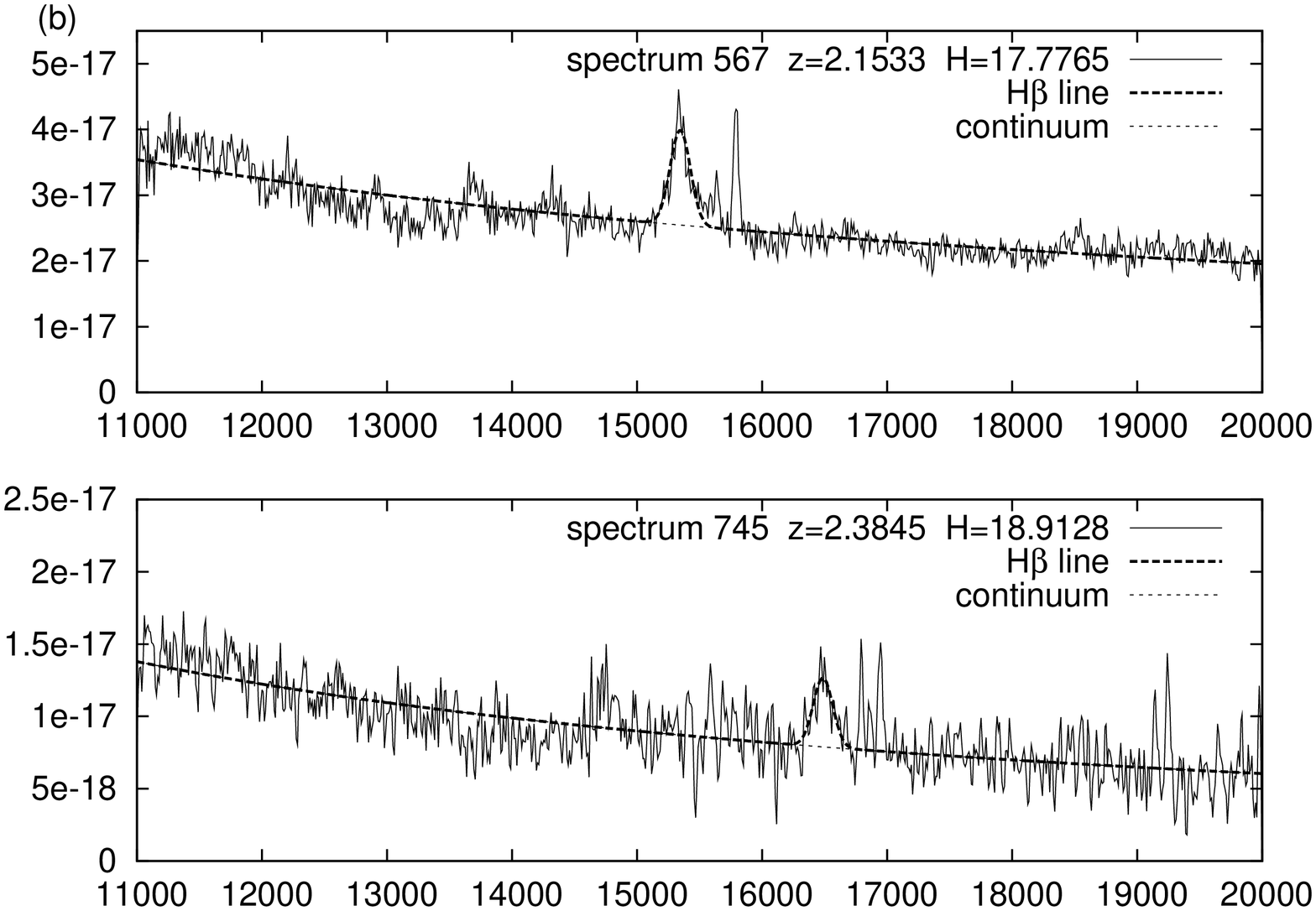}
\caption{Examples of IRAF Specfit fits to the spectral continuum and brightest broad line, plotted as flux $F_{\lambda}$, units ergs $\rm cm^{-2}s^{-1}\AA^{-1}$, against $\lambda$ in Angstroms. These are spectra with correct redshifts from but misclassification as emission-line galaxies. Two examples each are shown of fits to measure the FWHM of (a) $\rm H\alpha$ (b) $\rm H\beta$.} 
\end{figure}
\begin{figure} 
 \includegraphics[width=0.66\hsize,angle=-90]{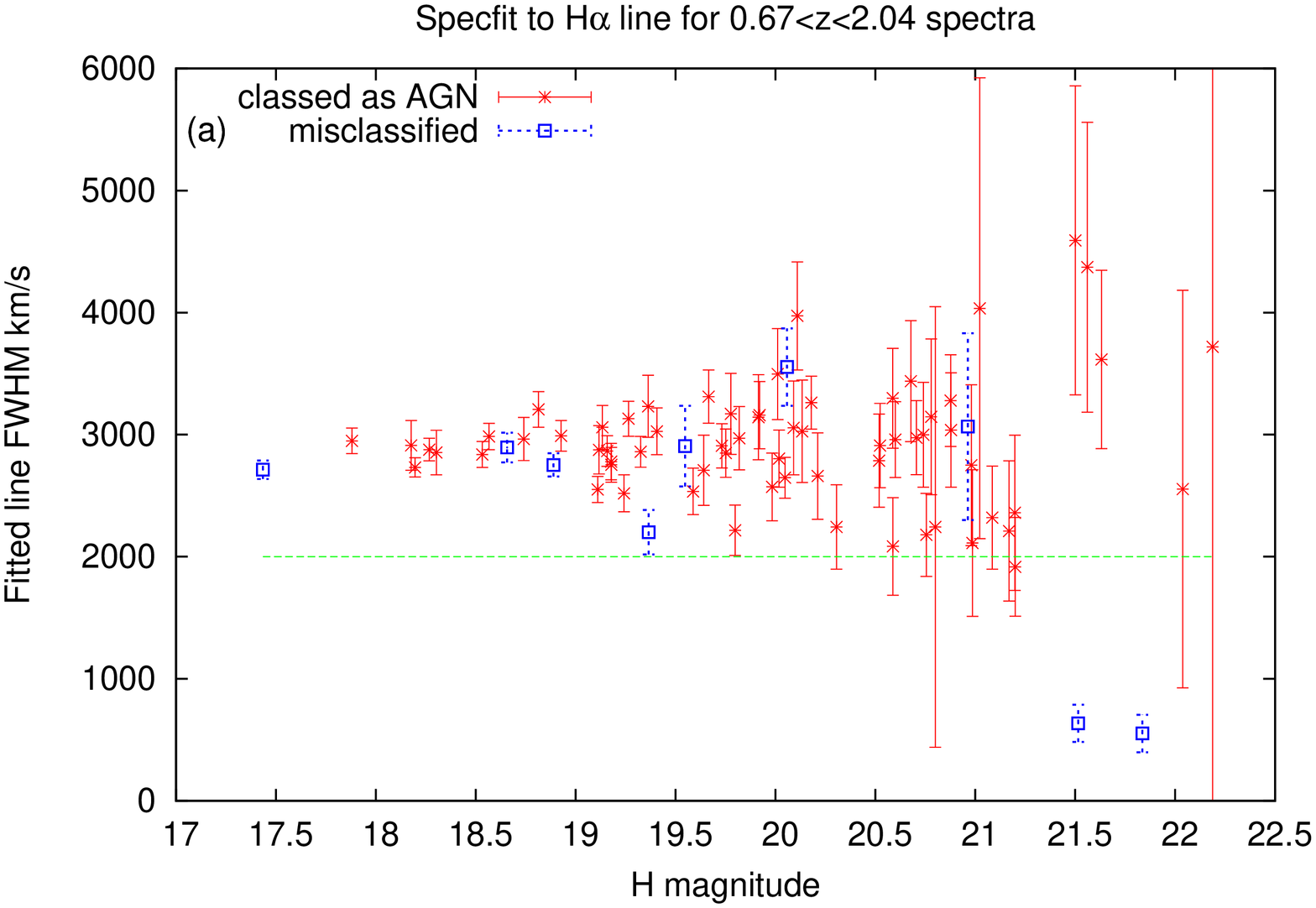}
\includegraphics[width=0.66\hsize,angle=-90]{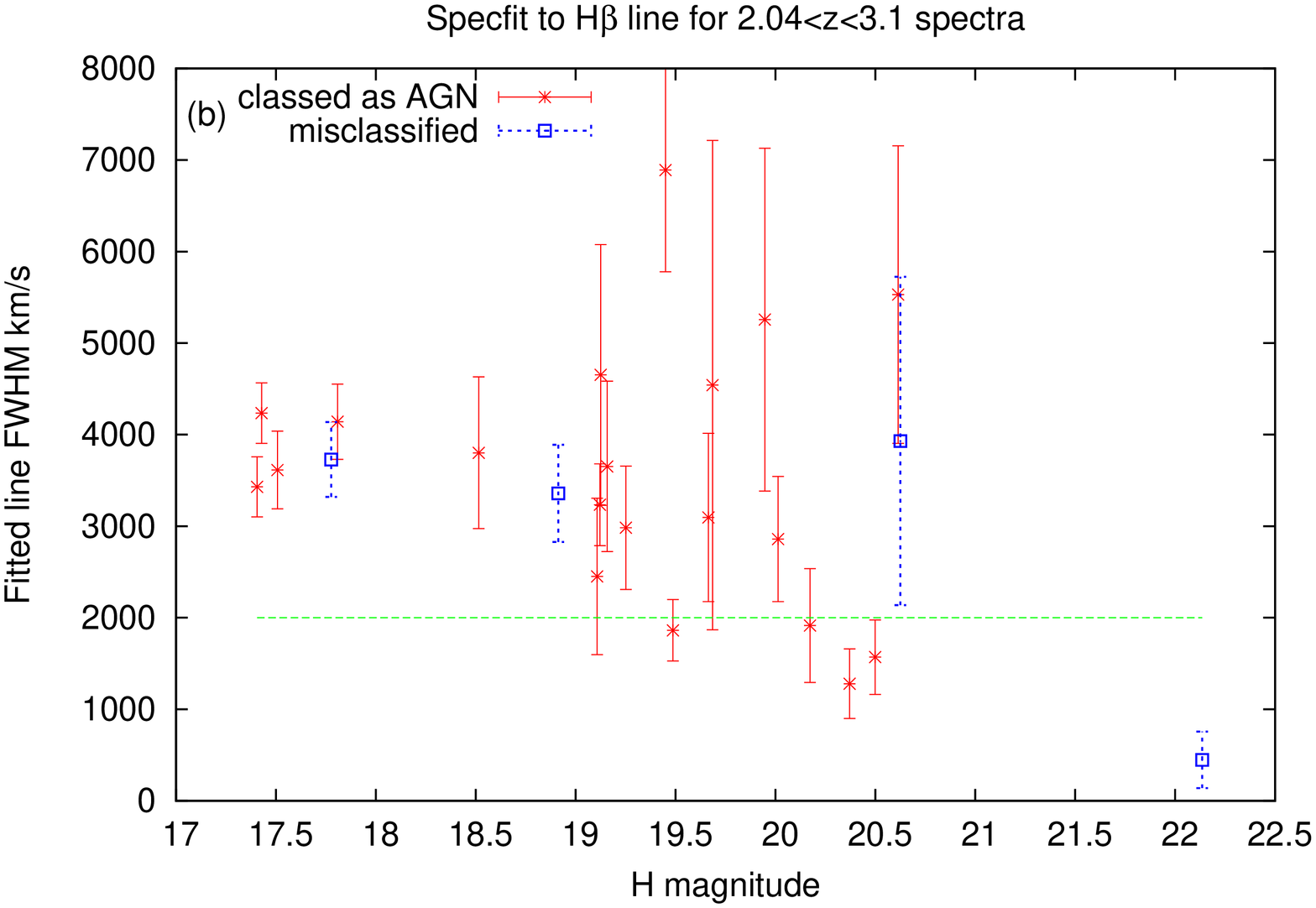}
\includegraphics[width=0.66\hsize,angle=-90]{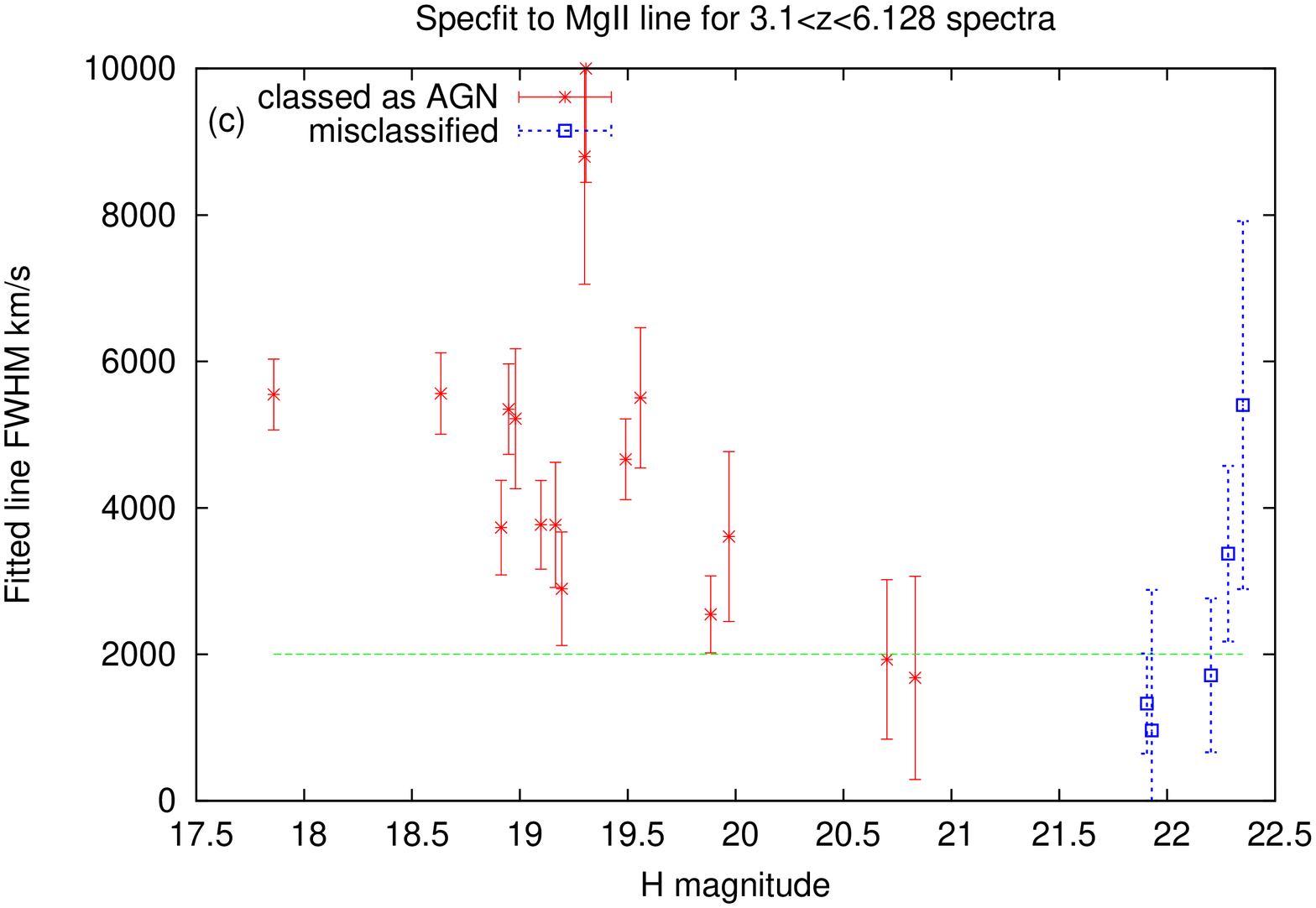}
\includegraphics[width=0.66\hsize,angle=-90]{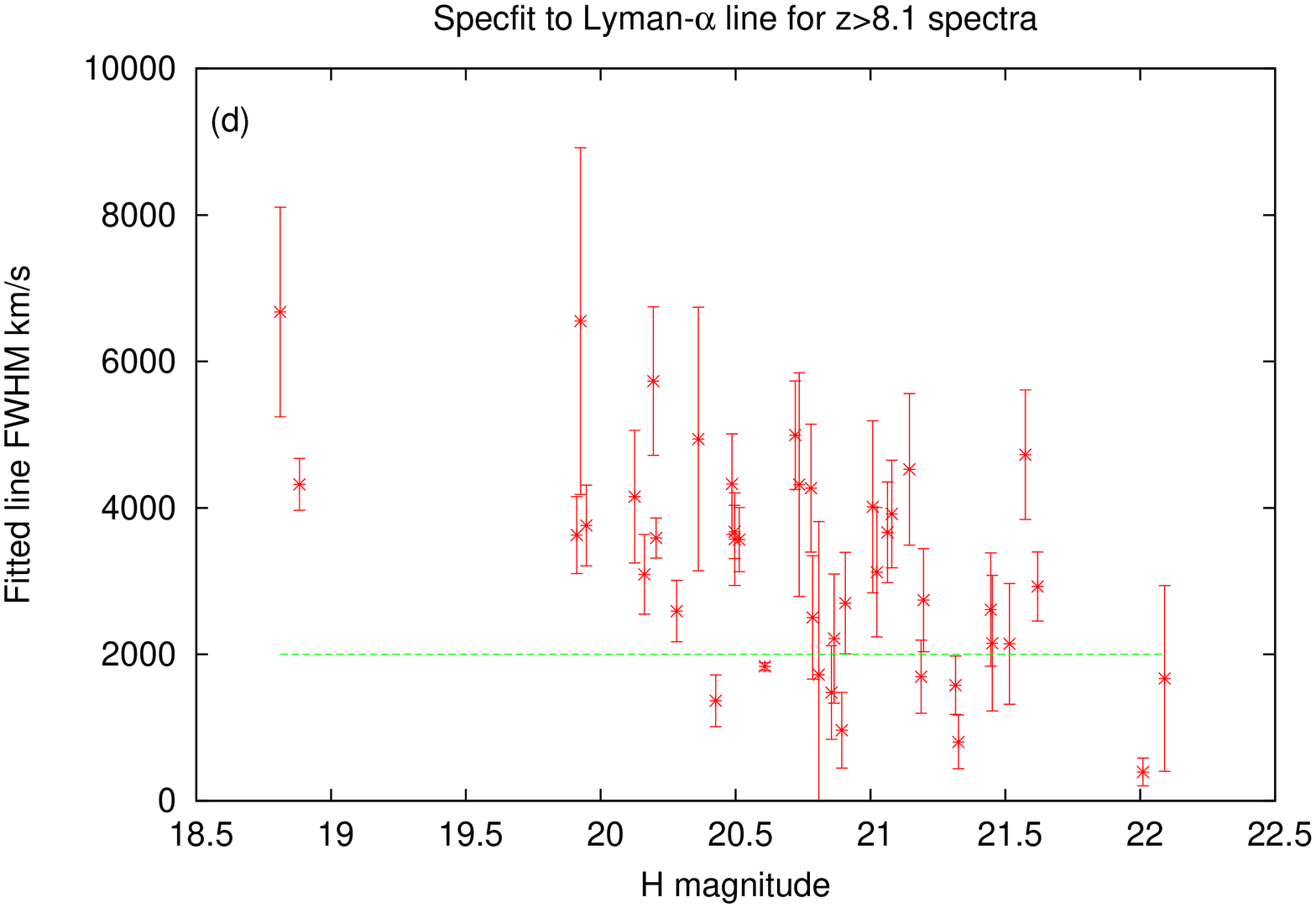}
\caption{Measurements of the FWHM of the brightest visible broad line, using IRAF Specfit, for spectra with correctly measured redshifts and both correct (AGN) and wrong classifications. Showing fits to (a) $\rm H\alpha$ (b) $\rm H\beta$ (c) MgII ($2800\rm \AA$) and (d) Lyman-$\alpha$.}
\end{figure}

For $\rm H\alpha$ this procedure works very well: line widths can be measured accurately to at least $H=20.3$, to which limit the mean FWHM is measured as 2923 km $\rm s^{-1}$ with scatter 318  km $\rm s^{-1}$. We can confidently reidentify as QSOs all 6 of the misclassified spectra in this redshift range and above this limit. For $\rm H\beta$ we get accurate FWHM to at least $H=19.3$, at which limit the mean FWHM is measured as 3607 km $\rm s^{-1}$ with scatter 587  km $\rm s^{-1}$, sufficient to reclassify two more of the spectra. For MgII the limit for good FWHM measurement may be shallower but is at least $H=19.1$ to  which the mean FWHM is measured as 4863 km $\rm s^{-1}$ with scatter 871  km $\rm s^{-1}$ (the only misclassified spectra here are  too faint to re-identify). In the Lyman-$\alpha$ window none of these spectra are misclassified but we can still test FWHM measurement. To $H=20.3$, the mean is 4410 km $\rm  s^{-1}$ but with a large scatter 1426 km $\rm s^{-1}$; a more sophisticated model may perform better due to the asymmetry and continuum break. 

By following redshift-finding with a measurement of line FWHM, we recover 8 more spectra into the `correct' list of QSO identifications, giving a total of 156.

\subsection{Lyman-$\alpha$ Identifications, and the Gains from Refitting Spectra}
\begin{figure} 
\includegraphics[width=0.7\hsize,angle=-90]{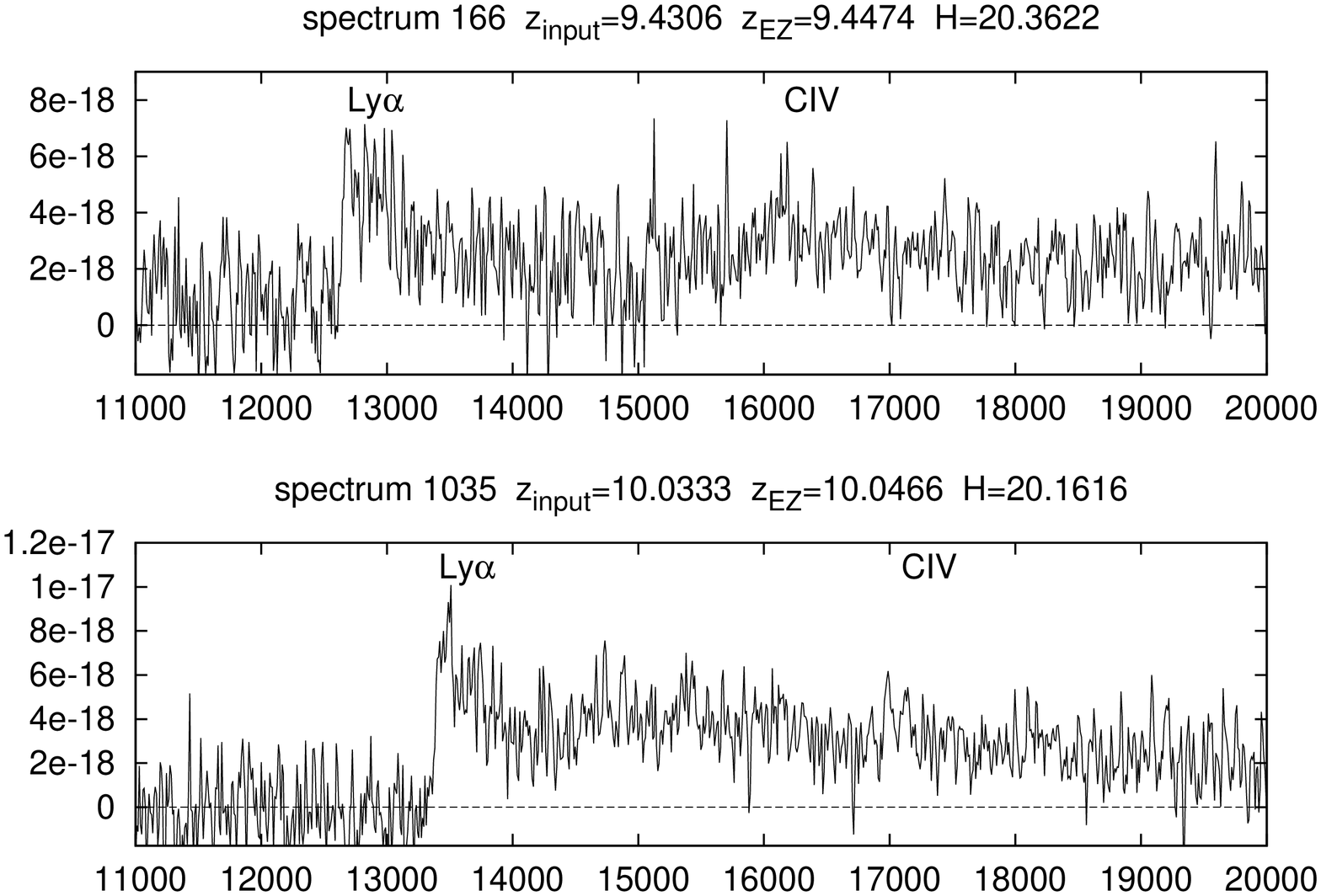}
\includegraphics[width=0.7\hsize,angle=-90]{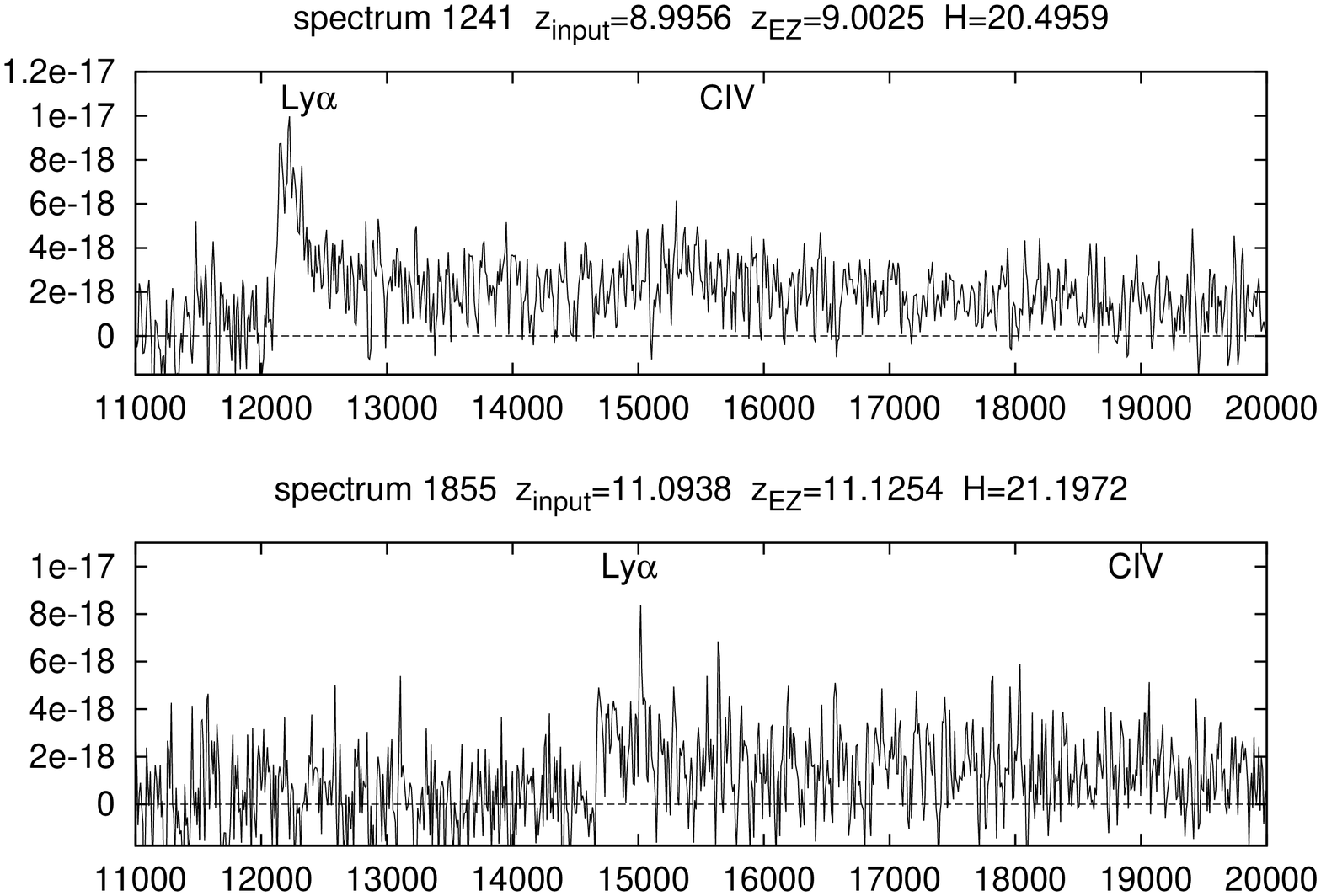}
\caption{Some fainter ($H>20$) examples of simulated spectra in the Lyman-$\alpha$ window for which the automated fit gave correct redshifts ($z_{EZ}$) and classifications.}
\end{figure}

\begin{figure} 
 \includegraphics[width=0.7\hsize,angle=-90]{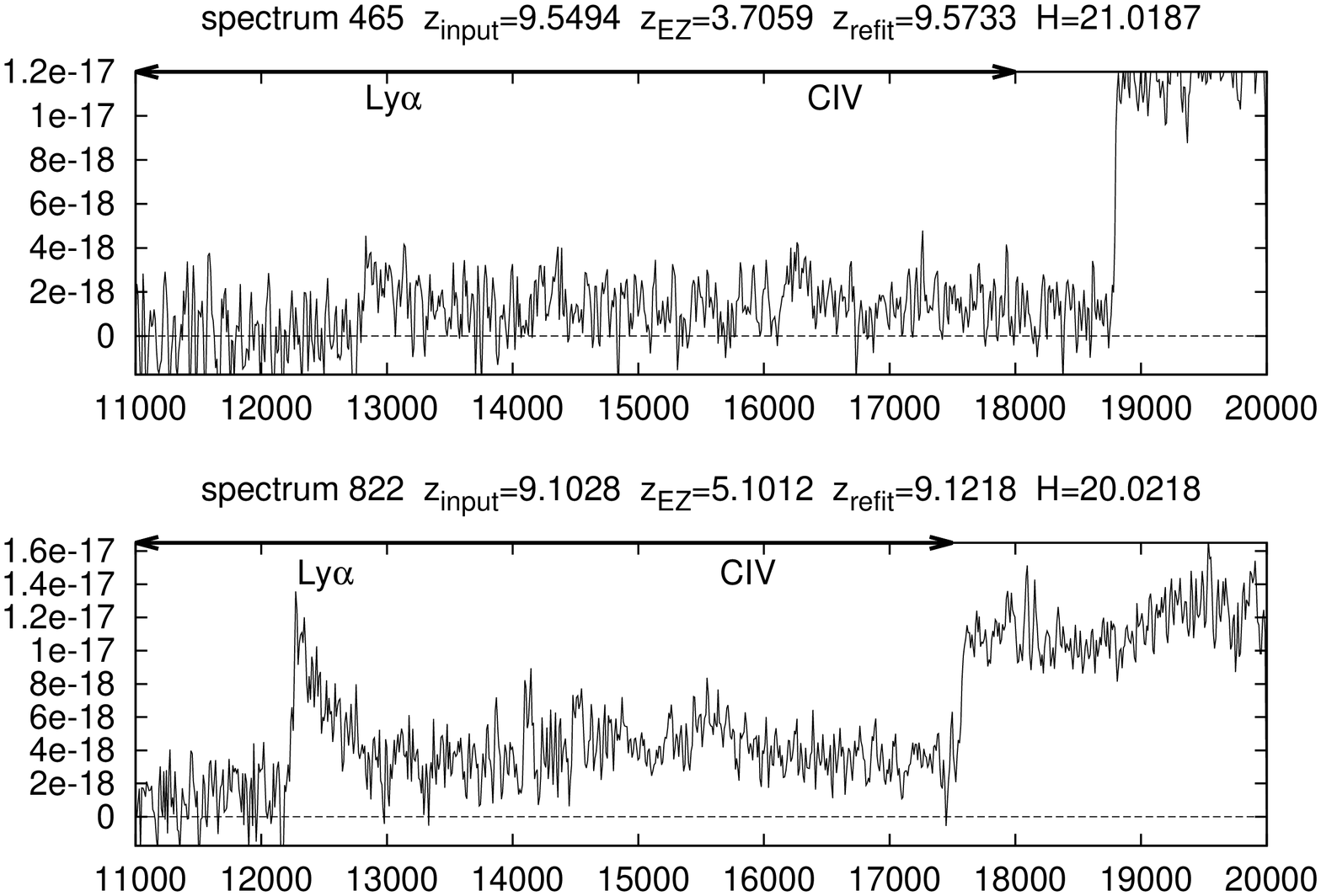}
\includegraphics[width=0.7\hsize,angle=-90]{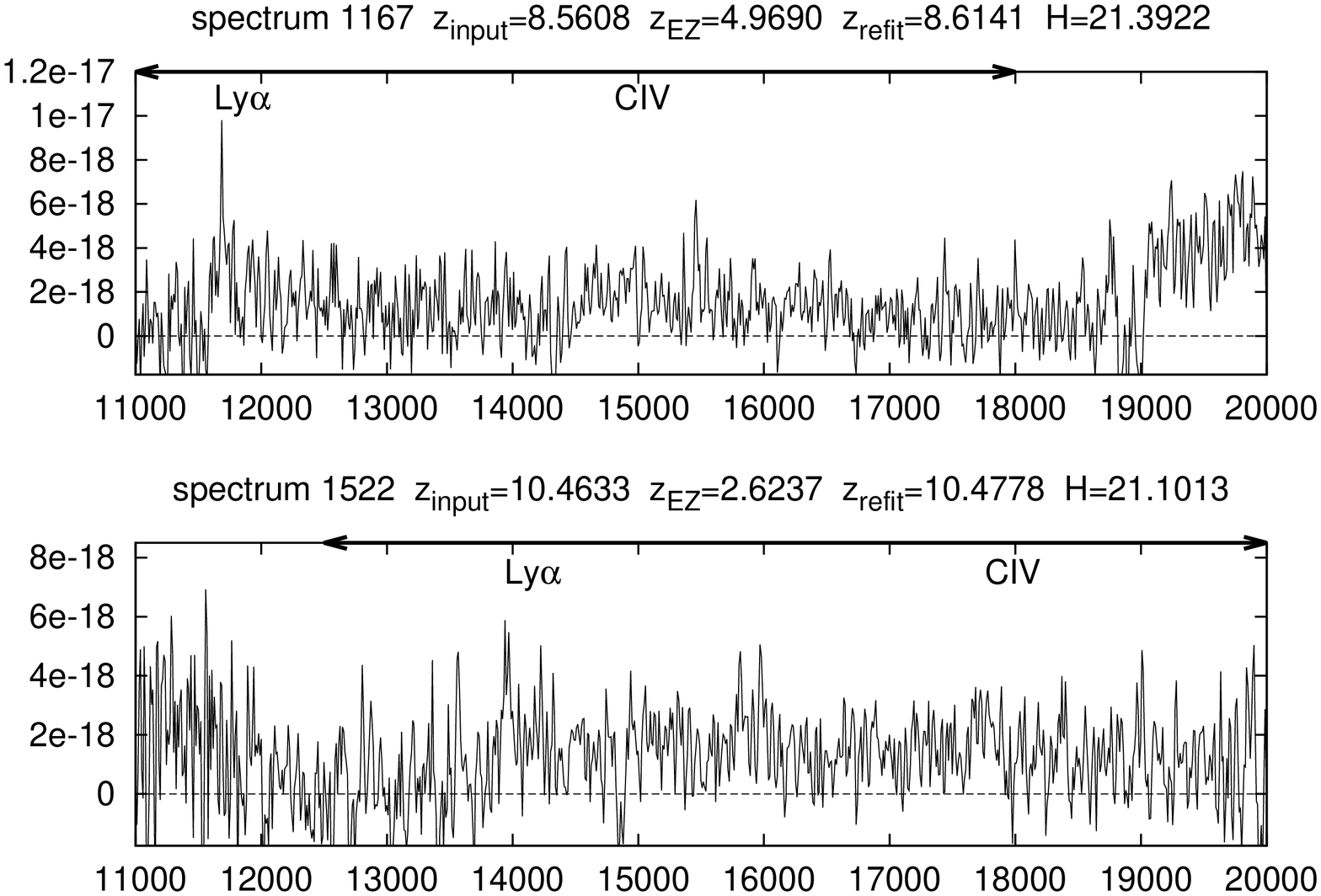}
\caption{Some examples of simulated spectra in the Lyman-$\alpha$ window, with contaminated regions (causing the discontinuities in the continuum level), for which the automated fit gave wrong redshifts ($z_{EZ}$), but subsequent re-fitting (restricting the wavelength range to that indicated by the arrowheads) returned the correct redshift ($z_{refit}$) and classification.}
\end{figure}

A key prediction of our simulation is that the ability to detect/recognize QSOs will greatly increase at the redshift where the strong Lyman-$\alpha$ line enters the spectrograph range ($z>8.06$ for {\it Euclid} NISP). In our simulation, {\sevensize EZ} in automated mode gave correct redshifts for 42 spectra at $z>8.06$ (examples on Fig. 12).  However, some fitted redshifts were wrong, even where the signal-to-noise should have been good enough and the Lyman-$\alpha$ was clearly visible, because of spectral contamination and overlapping. Often this affected only a fraction of the wavelength range, and if the bad region could simply be excluded from the fit, a correct redshift might be returned, but to do this the spectra have to be re-examined by eye.

Although it will be impractical or inefficient to examine by eye every spectrum, it will be possible to do this for the small subset of objects with evidence from multiband optical-NIR photometry that they are at $z>8$ ( i.e. they are Lyman-break objects which drop out of detection at $\lambda<1.0$--$1.1\rm \mu m$). Photometric selection could be from the imaging component of the {\it Euclid} wide survey, which should be able to select $z>8$ objects by $Y-J$ colour down to $J\simeq 22$ (Laureijs et al. 2011), and/or from ground-based optical-NIR surveys (UKIDSS, Vista Viking, Vista Hemisphere Survey etc.).

On this basis we re-examine by eye all of the $z>8.06$ spectra where the automated fit gave wrong redshifts, and where possible re-fit excluding the obviously contaminated wavelength range (other than for the $\lambda$ range, no further restrictions are placed on fitting, i.e the full range of redshifts and templates are available).

This method was very successful for spectra with a clearly visible Lyman-$\alpha$ line. We recovered correct redshifts and classification for 26 spectra at $z>8.06$, which previously had bad redshifts. This number includes 3 spectra where even after re-fitting $z_{EZ}$ is offset from $z_{input}$ by $0.05$--0.1; these were included on the grounds that Lyman-$\alpha$ is still correctly identified (a more sophisticated fitting method might improve the redshift accuracy). However, re-fitting was rarely successful (only for 2 spectra) for spectra with $H>21.5$ (generally too poor signal-to-noise ratio). Fig. 13 shows some examples of contaminated spectra successfully re-fitted in this way.

Objects at slightly lower redshifts ($7<z<8.06$) may also be selected by photometry, as $z-Y$ dropouts. We re-examined the brighter spectra in this redshift range, but in only 2 cases were able to obtain correct redshifts by refitting, and notably these were both `borderline' spectra at $8.02<z<8.06$, in which both the red wing of Lyman-$\alpha$ and the fainter CIV line were visible. 

Fig. 14 shows the positions of these 28 re-fitted spectra highlighted on the redshift-magnitude plot. If these are included the list of `correct' QSO identifications increases from 156 to 184.
 \section{QSO Detection rate as a function of magnitude and redshift}
 
  It is apparent from Fig. 14 that the limit to which QSO spectra can be correctly identified is very redshift-dependent, being deepest by far in the two windows where either $\rm H\alpha$ ($0.67<z<2.05$) or Lyman-$\alpha$ ($z>8.06$) is visible to NISP. In both these ranges there are many correct identifications at $H>21$, but at intermediate redshifts, there is none this faint. The intermediate range can be divided into two intervals, $2.05<z<3.0$ where $\rm H\beta$ and $\rm [OIII]$ are visible, and $3.0<z<8.06$ where there are only the `medium strength' broad-lines MgII, CIII] and/or CIV to provide a redshift; QSO identification is a little better in the first.
In the $\rm H\alpha$ and Lyman-$\alpha$ windows the detection rates fall to $e^{-1}$ (37\%) at $H\simeq 21.0$, while  for $2.05<z<3.0$ this occurs at $H\simeq 19.6$, and for $3.0<z<8.06$, $H\simeq 19.3$. In all ranges the detection rate does not cut off sharply, but declines over at least 1.5 magnitudes, and for real AGN, with large variations in line equivalent-widths, this fall-off would be even more gradual.

Dividing the spectra into these four redshift ranges, for each we calculate a detection rate as the proportion of (the extracted) spectra with both correctly (within $\pm 0.05$) measured redshifts and classification as QSO (including the 8 spectra reclassified using line FWHM), in $\Delta(H)=0.5$ mag intervals.
 A detection rate of unity is assumed brightward of the brightest simulated objects (e.g. $H<17$), which may not be quite correct, but very few high-$z$ QSOs will be this bright so this will not affect our conclusions.
 
 Fig. 15 shows the detection rates, and Table 2 gives the total ($N_{QSO}$, extracted spectra only) and correctly detected numbers ($N_{det}$, which does not include the 28 re-fitted spectra).
 If the 28 re-fitted $z>8$ spectra discussed above are included, this gives the detection rates plotted as bold lines in Fig. 15, and the detected numbers in columns $N_{detR}$ in Table 2. The detection rate in the Lyman-$\alpha$ window is substantially increased by re-fitting, especially at $21.0<H<21.5$ (from $25\%$ to $57\%$) although it still falls off steeply at $H>21.5$. With the addition of re-fitting we predict that the QSO detection rates will be higher and with a deeper limit (effectively at least $H\simeq 21.5$) for $z>8.06$ than in any other redshift range. 

\section{Predictions for QSO Detection in the {\it Euclid} NISP wide survey}
\subsection{How many QSOs will be found, including at $z>8$?}
Estimates can now be made of the total number of type 1 AGN that would be identified in the entire NISP wide survey (15000 $\rm deg^2$). Necessarily these will be  dependent on the evolution model. Fig. 16 shows QSO number counts from our adopted model (Bongiorno et al. 2007 with $k_{ev}=-0.47$ at $z>2.7$), now divided into 7  redshift intervals. These model counts are multiplied by our simulation's estimate of the QSO detection rate as a function of $H$ and $z$, to give the detected-QSO counts (Fig. 17; for the detection rates both without and with the inclusion of the re-fitted spectra). These detected counts, summed over $H$ for each redshift interval and multiplied by 15000 $\rm deg^2$, predict the total numbers that will be found in this area down to the faint limit of the survey (Table 3, columns 2 and 3). 

In total, we estimate the {\it Euclid} wide spectroscopic survey will find and identify about 1.41 million QSO, but with the great majority (92\%) in the $\rm H\alpha$ window. With $k_{ev}=-0.47$ evolution, the predicted number at $z>8.06$ (with automated redshift finding) is 22.27 (with the most distant one or two QSOs at $z=10$--12). With the improved rate of identification following by-eye refitting, this number increases by $59\%$ to 35.43.

Using instead the LF of Willott et al. (2010a), which is normalized at $z=6$ with evolution $k_{ev}=-0.47$ to $z>6$ (this cannot be extrapolated to lower redshifts) gives the predicted counts in columns 4 and 5 of Table 3. These are a factor 1.75 lower, with 12.68 in the Lyman-$\alpha$ window (increasing to 20.05 with refitting). As discussed in Section 2.3 our LDDE model gives a higher normalization at these redshifts (approximately that of the Jiang et al. 2009 LF). While at the time of writing the Willott et al. (2010a) LF is arguably the best estimate at $z\sim 6$, their Fig. 5 suggests that the true normalization for the bright end is probably intermediate between the two functions. On this basis, and as it is certain that photometric selection of $z>8$ objects will be available, we give a prediction for Lyman-$\alpha$ ($z>8.06$) QSOs detectable in the {\it Euclid} wide survey as 20--35.
\onecolumn
\begin{figure}
 \includegraphics[width=0.7\hsize,angle=-90]{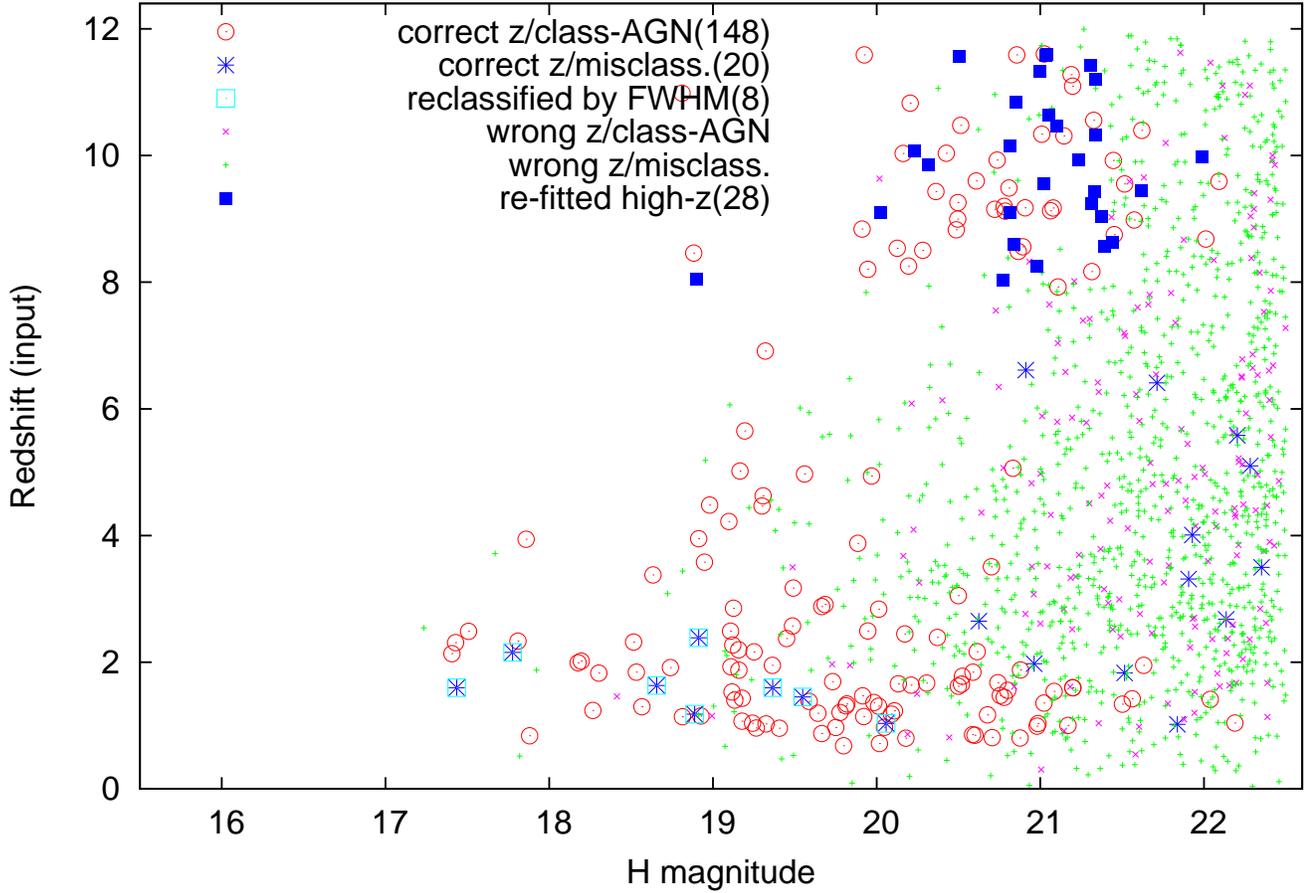}
\caption{Results of {\sevensize EZ} analysis of the simulated QSO spectra, shown on a plot of input redshift against H apparent magnitude. Symbols of 4 types indicate whether spectra have correct or incor- rect redshift measurements, and whether or not they have correct classifications (as QSOs). Additional symbols highlight the spectra (8) that were reclassified as AGN by line FWHM fitting, and the $z>8$ spectra with contamination, for which refitting `by-eye' gave correct redshifts (28).}
\end{figure}
\begin{table}
\begin{tabular}{lcccccccccc}
\hline
$H$ mag & \multispan{2} $0.67<z<2.05$ & \multispan{2}  $2.05<z<3$ &  \multispan{2} $3<z<8.06$ &  &  \multispan{2} $8.06 <z<12$ & \\
    & $N_{QSO}$ & $N_{det}$ & $N_{QSO}$ & $N_{det}$ & $N_{QSO}$ & $N_{det}$ & $N_{detR}$ & $N_{QSO}$ & $N_{det}$ & $N_{detR}$ \\
\hline
17.0--17.5 &  1 & 1 & 3 & 2 & 0 & 0 & 0 & 0 & 0 & 0\\
17.5--18.0 &  2 & 1 & 3 & 3 & 2 & 1 & 1 & 0 & 0 & 0\\
18.0--18.5 &  5 & 4 & 0 & 0 & 0 & 0 & 0 & 0 & 0 & 0\\
18.5--19.0 &  9 & 7 & 2 & 2 & 8 & 4 & 5 & 2 & 2 & 2\\
19.0--19.5 & 17 & 12 & 10 & 7 & 15 & 7 & 7 & 0 & 0 & 0\\
19.5--20.0 & 19 & 13 & 10 & 3 & 17 & 3 & 3 & 3 & 3 & 3\\
20.0--20.5 & 21 & 11 & 14 & 3 & 48 & 1 & 1 & 15 & 10 & 13\\
20.5--21.0 & 28 & 15 & 26 & 1 & 65 & 2 & 3 & 27 & 11 & 18\\
21.0--21.5 & 38 & 5 & 45 & 0 & 130 & 1 & 1 & 44 & 11 & 25\\
21.5--22.0 & 44 & 3 & 39 & 0 & 171 & 0 & 0 & 79 & 3 & 5\\
22.0--22.5 & 48 & 2 & 62 & 0 & 227 & 0 & 0 & 133 & 2 & 2\\
\hline
\end{tabular}
\caption{Number of QSO spectra included in the simulation $N_{QSO}$ (extracted spectra only) and the numbers with correct redshift and classification $N_{det}$, in $\Delta(H)=0.5$ mag bins of apparent magnitude and 4 intervals of redshift. The columns $N_{detR}$ (for the two highest redshift intervals only) show the (higher) numbers of detections
with the inclusion of the redshifts obtained using the method of refitting spectra (to exclude contaminated regions; see Section 4.4).}
\end{table}
\twocolumn

\begin{figure} 
 \includegraphics[width=0.7\hsize,angle=-90]{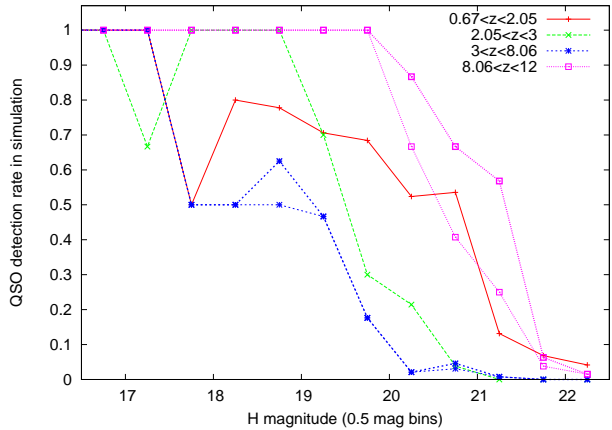}
\caption{The proportion of spectra correctly detected (in $\Delta(H)=0.5$ magnitude bins, divided into four intervals of redshift. Detection limits are  deepest in the `windows' where $\rm H\alpha$ ($0.67<z<2.05$) or Lyman-$\alpha$ ($8.06<z<12$) are visible. For the highest 2 redshift intervals the detection rates are plotted without and with (bolder lines, same colours) the inclusion of the 28 refitted redshifts. }
\end{figure}
\begin{figure}
\includegraphics[width=0.7\hsize,angle=-90]{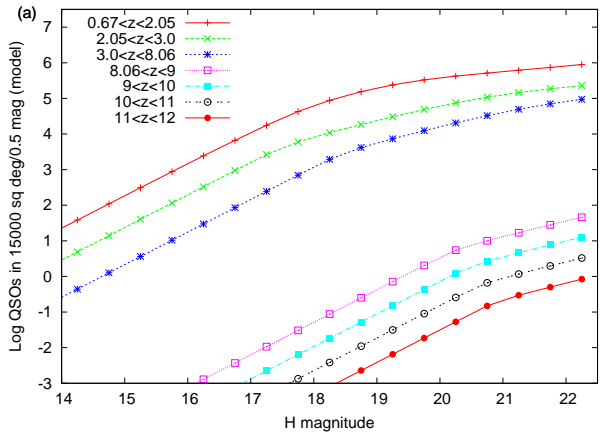}
\caption{$H$-band differential QSO number counts from our model, divided by redshift.}
\end{figure}
\begin{figure}
\includegraphics[width=0.7\hsize,angle=-90]{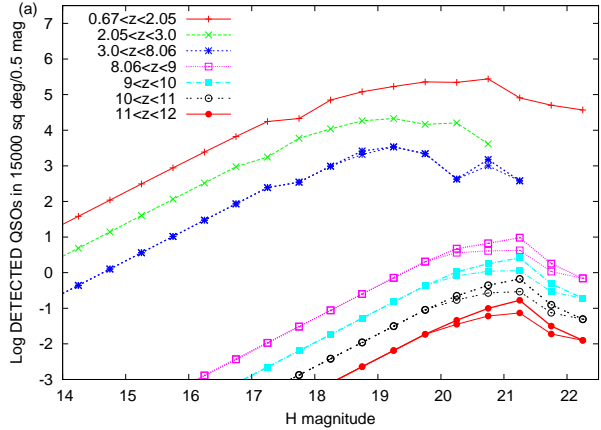}
\caption{The number counts from the model multiplied by the simulation's detection rate (as a function of $H$ and $z$), calculated excluding and (as bold lines) including refits, to predict the count of correctly identified QSOs.}
\end{figure}

However, it has been suggested (Steve Warren, private communication) on the basis that UKIDSS has so far found only 1 QSO at $z>6.5$, that the negative $\Phi^*$ evolution at these redshifts could be steeper with $k_{ev}\simeq -0.72$ (extrapolated from a Willott et al. LF at $z=6$). Although current observations remain consistent with $k_{ev}=-0.47$, there are models (e.g. Shankar et al. 2010; DeGraf et al. 2011) that predict a faster evolution at $z>6$, approximately in this $k_{ev}$ range. With the Willott et al. LF and  $k_{ev}=-0.72$, the expected count at $z>8.06$ would be reduced by more than a factor 4, to 2.81 (4.43 with refitting), with very little chance of finding a $z>10$ QSO. 

Whatever the evolution rate, the estimates in given in Table 3 should probably be subject to a downward correction of about $6\%$ because of the small fraction of very weak-lined or lineless QSOs (see Section 2.1), which with the methods of redshift finding discussed in this paper  would probably be missed. The prediction for Lyman-$\alpha$ detections is then reduced to 19--33.

It might also be possible that at $z>8$, the fraction of intrinsically weak-lined/lineless  QSOs could increase significantly above that seen at $4<z<5$, and/or that for $z>8$ QSOs the blocking of Lyman-$\alpha$ by HI in the IGM could be much greater than assumed in our model (e.g. the ionised zones are much smaller). If either of these occur it would lower the number of detections to some (unknown) extent.
\begin{table}
\begin{tabular}{lccccc}
\hline
$z$ range & \multispan{2}\hfil Our model LF \hfil &\multispan{2}\hfil  W10 LF \hfil\\
   & no refit & refit & no refit & refit  \\
\hline

$0.67<z<2.05$ & 1301429. &   1301429. & - & - \\
$2.05<z<3.0$  & 94538. & 94538. & - & - \\
$3.0<z<8.06$  & 11193. & 12215. & - & - \\
$8.06<z<9.0$  & 16.85 &  26.62 & 9.60 & 15.16 \\
$9.0<z<10.0$  & 4.19  &  6.78 &  2.38 & 3.78\\
$10.0<z<11.0$ & 0.99  & 1.64 & 0.57 & 0.90\\
$11.0<z<12.0$ & 0.23  & 0.39 & 0.13 & 0.21 \\
\hline
\end{tabular}
\caption{The total number of QSOs in different redshift ranges detectable in a 15000 $\rm deg^2$ wide {\it Euclid} NISP survey, for this paper's evolution model (with $k_{ev}=-0.47$ negative density evolution at all $z>2.7$), and (at $z>6$ only) the Willott et al. (2010) QSO luminosity function (W10) with $k_{ev}=-0.47$ evolution at $z>6$. Shown for the detection rates estimated without and with re-fitting contaminated high-$z$ spectra.}
\end{table}
By the time {\it Euclid} is launched, the normalization of the QSO LF at $z=6$ should be very well determined, and there may be further detections at $z\simeq 7$, so the numbers  in the NISP Lyman-$\alpha$ window will constrain the evolution at $z\geq 7$. With careful analysis of the highest redshift spectra (such as the refitting by eye tried here) to maximize detections, the survey should have good enough statistics to distinguish (at high significance) between a continuation of $k_{ev}\simeq -0.47$, and the $k_{ev}\simeq -0.72$ or steeper evolution of the maximally rapid (unity duty cycle Eddington-rate) accretion models. In addition to giving an estimate of $k_{ev}$, the redshift distribution of the detections will directly trace the evolution over $8<z<10$--12.

\subsection{X-ray Fluxes of $z>8$ QSOs}
It is of interest whether $z>8$ QSOs will also be detected in forthcoming wide X-ray surveys. QSO X-ray luminosities can be estimated approximately from optical luminosities, $M_B$.
Fig. 18 shows the $B$-band absolute magnitudes of the 184 simulated QSO spectra that would be detected (including the re-fits). The QSOs detectable (with {\it Euclid}) in the Lyman-$\alpha$ window have luminosities ranging from $M_B=-25.87$ up to $M_B=-29.34$ with most at 
 $-28<M_B<-26$. All are above $M^*$ ($M_B\simeq -25.81$) and thus lie on the bright-end slope, in the same range of $M_B$ as the highest redshift ($z\sim 6$) QSOs found in the SDSS (e.g. Jiang et al. 2009) and the $z=7.085$ discovery of Mortlock et al. (2011), for which $M_B\simeq -27.2$. Note that {\it Euclid} could detect $z>8$ QSOs at least $\sim 1$ magnitude less luminous than ULAS J1120+0641.

\begin{figure}
\includegraphics[width=0.7\hsize,angle=-90]{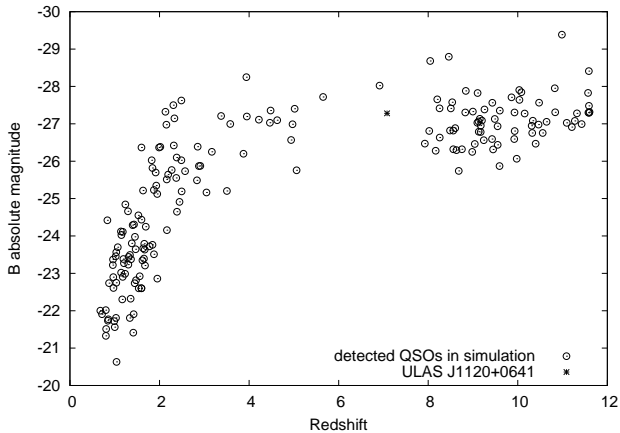}
\caption{The $B$-band absolute magnitudes for the 184 QSO spectra in our simulation which have correct redshifts (including refits) and classification. Most in the Lyman-$\rm \alpha$ window have $-28<M_B<-26$.}
\end{figure}

 Lusso et al. (2010) investigate the UV to X-ray ratio for a
large XMM-COSMOS sample of type 1 broad-line AGN, with a very wide
redshift (up to $z=4$ with $z_{mean}=1.66$) and luminosity range.
They fit a non-linear relation 

\smallskip
${\rm log}~L(2~{\rm keV})=0.760~{\rm log}~L(2500 $\rm \AA$) +3.508 $
\smallskip

\noindent Assuming, from the SDSS QSO template, $F_\nu(2500{\rm \AA})=0.8789 F_\nu(B)$,
and from Tozzi et al. (2006), that typical QSO spectrum at 0.5--7 keV is a power law
$F_\nu \propto \nu^{-0.8}$ (photon index 1.8), we derive 

\smallskip
log $L(0.5-2.0~{\rm keV})=36.917 -0.304 M_B$ in ergs $\rm s^{-1}$
\smallskip

Fluxes can be estimated from this luminosity, with the k-correction in log flux
being $0.2~{\rm log}~(1+z)$.

Fig. 19 shows the predicted $F(0.5-2.0 {\rm keV})$ fluxes (observer-frame) for the simulation's detected QSOs, compared to $5\sigma$ point-source detection limits in this band for the forthcoming eROSITA (extended Roentgen survey with an imaging telescope array) all-sky survey and deep fields (Capelluti et al. 2010), and for the all-sky and medium and deep surveys with the proposed Wide Field X-ray Telescope, WFXT (Brusa et al. 2010; Gilli et al. 2010). These 4 limits are respectively $1\times 10^{-14}$, $2\times 10^{-15}$, $3\times 10^{-15}$,  $5\times 10^{-16}$ ergs $\rm cm^{-2}s^{-1}$.

It is predicted that $z>8.06$ QSOs detectable with {\it Euclid} would generally have 0.5--2.0 keV fluxes 1--$4\times 10^{-15}$ ergs $\rm cm^{-2}s^{-1}$. None of these (unless they have unusually high X-ray/optical ratios) would  be detectable in the eROSITA all-sky survey. Some may be within the limit for the eROSITA deep fields at the ecliptic poles, which cover much smaller areas ($\sim 200$ $\rm deg^2$ total) but by chance might contain one.

If the more sensitive WFXT is launched and attains its proposed limits, Fig. 19 suggests about ${1\over 7}$  of {\it Euclid}-NISP's $z>8.06$ QSO detections will be within reach of its all-sky survey. With $k_{ev}\simeq -0.47$ evolution this sample would in total be 20--35, so the {\it Euclid}-WFXT overlaps would be 3--5 QSOs. All {\it Euclid}-detectable QSOs should easily within the limits for the WFXT medium survey, which is planned to cover a large area of 3000 $\rm deg^2$ and so would include and detect a further 4--7 of the {\it Euclid} $z>8.06$ QSOs. It will probably detect further $z>8$ QSOs too faint for {\it Euclid}  spectroscopy, but for which {\it Euclid} imaging would provide photometric evidence for $z>8$. We conclude {\it Euclid} would be well matched with WFXT for studying the highest redshift ($z>8$) AGN (eROSITA appears not to be sensitive enough).

\begin{figure}
\includegraphics[width=0.7\hsize,angle=-90]{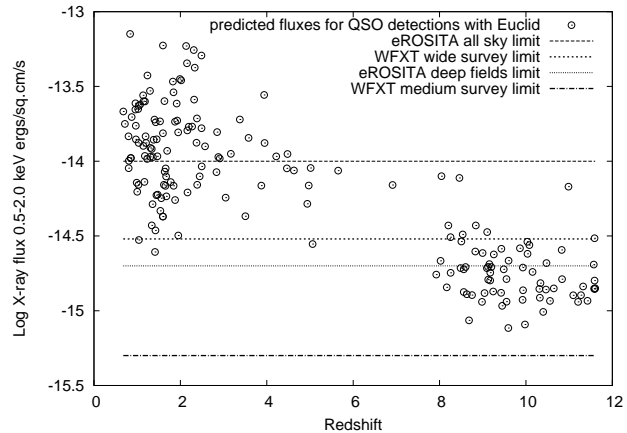}
\caption{The 0.5--2.0 keV X-ray fluxes predicted (from the $M_B$ magnitudes) for all (182) correctly identified QSOs in the simulation, compared with expected detection limits for the eROSITA all-sky survey and deep fields, and the wide and medium survey with the proposed WFXT.}
\end{figure}
\section{Summary and Discussion}

(i) We simulate the detectability of type 1 QSOs in a wide-field near-infra-red slitless spectroscopic survey, using the parameters of the {\it Euclid} near-infrared spectrometer and imaging photometer (NISP). The spectra are represented using the SDSS composite QSO template of Vanden Berk et al. (2001), extended shortward and longward following Telfer et al. (2002) and Richards et al. (2006) respectively, and the simulation included QSOs at all redshifts out to $z=12$. We add the effect of the (redshift-dependent) IGM absorption on the spectrum at $\lambda_{rest}<1216\rm \AA$, using the model of Madau (1995).

At the redshifts ($z>8.06$) where the strong  Lyman-$\alpha$ emission line enters the  the NISP wavelength range (11000--$20000\rm \AA$), the absorption from HI will be very strong, consequently  Lyman-$\alpha$ in these spectra will appear asymmetric because of absorption on the blue wing and there will be very sharp and deep spectra break at $\lambda_{rest}\simeq \rm 1200\AA$. However, due to the `bubble' of ionized hydrogen surrounding a high-$z$ QSO (which is also represented in our model) the peak of the Lyman-$\alpha$ line should not be absorbed at all and will remain visible.
\medskip

(ii) In the simulation the QSO luminosity function and its evolution were modelled using the luminosity dependent density evolution model of Bongiorno et al. (2007). At $z>2.7$ this was extrapolated by a simple exponential decline in density, $\phi\propto 10^{-0.47(z-2.7)}$ ($k_{ev}=-0.47$). As the proportion of very high redshift QSOs in any flux-limited sample will be extremely small, their numbers are scaled up by a large factor ($10^{0.58z}$) in the simulation, while retaining the magnitude and redshifts from the model. The QSO simulation covers a $0.5\times 0.5$ $\rm deg^{2}$ area and is limited at $H=22.5$.
The effects of spectral contamination and overlap are represented in part by the artificially high density of QSO spectra but also by mixing the QSOs with faint emission-line galaxy spectra.
\medskip

(iii) The 1468 QSO spectra that could be extracted from the {\sevensize AXESIM} image were analysed with an automated redshift finder ({\sevensize EZ}), which initially found correct redshifts (within $\Delta(z)=\pm 0.05$) and classifications (as QSO spectra) for 148. For a further 20 the redshifts were correct but the spectra were misidentified as non-AGN galaxies. The second step of our analysis was to fit the brightest lines with Gaussian profiles, and if the line $\rm FWHM\geq 2000$ km $\rm s^{-1}$, the spectrum (if initially classed as a galaxy), is reclassified as type 1 AGN. Using {\sevensize IRAF} `specfit', line FWHM could be measured well enough to confidently identify type 1 AGN to at least $H\simeq 20.3$ where $\rm H\alpha$ is visible and $H=19.3$ using $\rm H\beta$, and we could reclassify 8 of the 20 misclassified spectra as QSOs.

The successful detection rate (i.e. fraction of spectra with both correct redshift and classification) is highest, and detection reached deeper limits
($H\simeq 21$--21.5) in the two redshift windows where either $\rm H\alpha$ ($0.67<z<2.05$) or Lyman-$\alpha$ ($z>8.06$) is visible in the spectrograph $\lambda$ range (this is not surprising as these are by far the strongest lines in terms of flux and equivalent width). Detection limits are less deep where the redshift is provided by $\rm H\beta$ and $\rm [OIII]$ ($2.05<z<3.0$) and shallowest in the intermediate range ($3<z<8.06$) where the useful broad lines are MgII, CIII] and CIV.

\medskip

(iv) In some cases the automated fit gave a wrong redshift because of contamination affecting part (even a small part) of the spectrum.
However, {\it Euclid} and/or ground-based optical--NIR imaging will provide an independent photometric selection  of $z>8$ objects within the limits of the spectroscopic survey, e.g on the basis of a drop-out in $Y-J$. All spectra flagged in this way could then be carefully examined by eye and re-fitted, with exclusion from the fit of obviously contaminated wavelength regions (generally either at the blue or the red end). In this way it is possible to quickly and very significantly improve the QSO detection rate at $z>8$.

We performed  new fits on the $z>8$ spectra where automated fitting had given wrong redshift measurements, restricting the fit's wavelength range (by eye) to exclude obviously contaminated regions while including the Lyman-$\alpha$ line. In this way we
obtained the correct redshifts for a further 28 spectra, almost all at $z>8.06$ and $H<21.5$.  
For QSOs in the range $21<H<21.5$, the inclusion of refit redshifts  gave more than a doubling of the estimated QSO detection rate (from 25\% to 57\%). This method did not work well at $H>21.5$ (signal-to-noise ratio too low) or at $7<z<8$ (Lyman-$\alpha$ not visible). Over all magnitudes, by-eye refitting would increase the sample of detected  $z>8.06$ QSOs by a very substantial $\sim 60\%$. With refitting, {\it Euclid} would now attain higher detection rates and deeper limits (effectively $H=21.5$) for QSOs at $z>8.06$ than in any other redshift range, thanks to the strength of the Lyman-$\alpha$ line and break. 
\medskip

(v) Combining the QSO detection rate from our simulation with the LDDE model for QSOs, we predict that in the {\it Euclid} NISP wide survey, area 15000 $\rm deg^2$, a total of 1.41 million type 1 QSOs will be found, with most ($92\%$) in the $\rm H\alpha$ window. 
In the Lyman-$\alpha$ window ($z>8.06$), with the LDDE model and $k_{ev}=-0.47$ evolution we predict 22.6 QSOs will be detected and identified. Using instead the Willott et al. (2010a) LF for $z\sim 6$ QSOs, again with $k_{ev}=-0.47$ evolution, this falls to 12.7. We took these two estimates as spanning the probable range of uncertainty in the (bright-end normalisation of the) QSO luminosity function at $z\sim 6$. With refitting contaminated spectra, {\it Euclid}'s 
detection rates at these redshifts are $\sim 60\%$ higher, and the range of prediction for $z>8.06$ QSOs increases from 12.7--22.6 to the range 20--35. This reduces again slightly to 19--33 if a small down-correction is applied for the 6\% fraction (from Diamond-Stanic et al. 2009) of very weak-lined and lineless QSOs.
Out of this number, we predict {\it Euclid}'s highest redshift detection is likely to be at $z\sim 10$--11. 

However, the numbers of QSO detections at  $z>8.06$ would reduce to only 4 or 5 if the evolution at $z>6$ steepens to $k_{ev}\simeq -0.72$, which is approximately as predicted by models with Eddington-rate accretion and a duty cycle of unity (e.g. Shankar et al. 2010). The {\it Euclid} wide survey will give good enough statistics on $z>8$ QSOs to distinguish between these two possibilities, estimate $k_{ev}$ and set important constraints on the early evolution of AGN.

Any QSOs detectable at $z>8.06$ would lie on the bright-end slope of the AGN luminosity function ($-26>M_B>-28$), with SMBH masses ranging from at least $\sim 10^{8.5} M_{\odot}$, on the basis of, e.g. the $M_{BH}$ estimates of Willott et al. (2010b), up to the $\sim 2\times  10^9M_{\odot}$ of  ULAS J1120+0641. Their discovery at $z>8$ epochs would further challenge models to explain their early/rapid formation and may favour larger masses for the seed black holes (e.g. $\sim 10^5 M_{\odot}$ as in the nuclear star cluster collapse scenario of Davies, Miller and Belovary 2011). 
\medskip

(vi) Using the QSO X-ray to UV relation of Lusso et al. (2010), we estimate the X-ray fluxes of the {\it Euclid}-detectable $z>8$ QSOs would typically be $F(0.5$--$2.0 {\rm keV})=1$--$4\times 10^{-15}$ ergs $\rm cm^{-2}s^{-1}$, too faint for the eROSITA all-sky survey to detect. However, some of these AGN would be detectable on the proposed WFXT wide survey, and all would be within the flux limits of the WFXT medium survey. With $k_{ev}=-0.47$ evolution we predict a $z>8.06$ overlap between {\it Euclid} wide and  WFXT wide of 3--5 QSOs, and of 4--7 QSOs between {\it Euclid} wide and the 3000 $\rm deg^2$ WFXT medium survey; evidently the two telescopes are well matched for the highest redshift QSOs. 
\medskip

\section*{Acknowledgements}
The authors acknowledge the support from the Agenzia Spaziale Italiana
(ASI-Uni Bologna-Astronomy Dept. `{\it Euclid}-NIS' I/039/10/0), from
MIUR PRIN 2008 `Dark energy and cosmology with large galaxy surveys, 
and from ASI/INAF contract I/009/10/0.

\section*{References} 

\vskip0.15cm \noindent Becker R.H, et al. 2001, AJ 122, 2850.

\vskip0.15cm \noindent Bongiorno A. et al., 2007, A\&A, 472, 443.

\vskip0.15cm \noindent Bolton J.S, Haehnelt M.G., Warren S.J., Hewett P.C., 
Mortlock D.J., Venemans B.P., McMahon R.G., Simpson C., 2011, MNRAS 416, 70.

\vskip0.15cm \noindent Boyle B.J., Shanks T., Croom S.M., Smith R.J., Miller L., Loaring N., Heymans C., 2000, MNRAS, 317, 1014.

\vskip0.15cm \noindent Brusa M. et al. 2009, ApJ, 693, 8. 

\vskip0.15cm \noindent Brusa M., Gilli R., Civano F., Comastri A., Fiore F., Vignali C.,
2011, Mem. Soc. Astron. Ital. Suppl., 17, 106.

 \vskip0.15cm \noindent Cappelluti N.,et al. 2011, Mem. Soc. Astron. Ital. Suppl., 17, 159.

\vskip0.15cm \noindent Carilli C.L., et al., 2010, ApJ, 714, 834.

\vskip0.15cm \noindent Civano F., et al. 2011, ApJ, 741, 91.

\vskip0.15cm \noindent Croom S.M., et al. 2009, MNRAS 399, 1755. 

\vskip0.15cm \noindent Davies M.B., Miller M.C., Bellovary J.M., 2011, ApJ, 740, 42.

\vskip0.15cm \noindent DeGraf C., Di Matteo T., Khandai N., Croft R., Lopez J., 
Springel V., 2011, MNRAS, submitted (astro-ph/1107.1254).

\vskip0.15cm \noindent Diamond-Stanic A.M., et al., 2009, ApJ 699, 782.

\vskip0.15cm \noindent Fan X., et al., 1999, ApJ 526, L57.

\vskip0.15cm \noindent Fan X., et al., 2003, AJ 125, 1649.

\vskip0.15cm \noindent Fan X., et al., 2004, AJ 128, 515.

\vskip0.15cm \noindent Fan X., et al., 2006, AJ 131, 1203.

\vskip0.15cm \noindent Fontanot F., Cristiani S., Monaco P., Nonino M., Vanzella E., 
Brandt W.N., Grazian A., Mao J., 2007a, A\&A 461, 39.

\vskip0.15cm \noindent Fontanot F., Somerville R.S., Jester S., 2007b, preprint( arXiv:0711.1440) (F07b).

\vskip0.15cm \noindent Garilli, B., Fumana, M., Franzetti, P., Paioro L., Scodeggio M., Le F{\`e}vre O., Paltani S., Scaramella R., 2010, PASP, 122, 827.

\vskip0.15cm \noindent Gilli, R. 2011, Mem. Soc. Astron. Ital. Suppl., 17, 85.

\vskip0.15cm \noindent Goto T., Utsumi Y., Furusawa H., Miyazaki S., Komiyama Y., 2009, MNRAS 400, 843.

\vskip0.15cm \noindent Hao Lei, et al., 2005, AJ 129, 1783.

\vskip0.15cm \noindent Hutchings J.B., Frenette D., Hanisch R., Mo J., Dumont P.J., Redding D.C., Neff S.G, 2002, AJ 123, 2936.

\vskip0.15cm \noindent Jiang L., Fan X., Vestergaard M., Kurk J.D., Walter F., Kelly B.C., Strauss M.A., 2007, AJ 134, 1150.

\vskip0.15cm \noindent Jiang L., et al., 2008, AJ, 135, 1057.

\vskip0.15cm \noindent Jiang L., et al., 2009, AJ, 138, 306.

\vskip0.15cm \noindent K{\"u}mmel M., Walsh J. R., Pirzkal N., 
Kuntschner H., Pasquali A., 2009, PASP, 121, 59

\vskip0.15cm \noindent Kurk J.D., Walter F., Fan X., Jiang L., Jester S., Rix H.-W.,
Riechers D.A., ApJ, 702, 833.
 
\vskip0.15cm \noindent Laureijs R., et al., 2011, eds.,  {\it Euclid}: Mapping the Geometry of the Dark Universe. Definition Study Report, European Space Agency, Noordwijk, Netherlands (arXiv:1110:3193).

\vskip0.15cm \noindent Lehnert M.D., et al., 2010, Nature 467, 940.

\vskip0.15cm \noindent Lusso E., et al., 2010, A\&A, 512, 34.

\vskip0.15cm \noindent Madau P., 1995, ApJ 441, 18.

\vskip0.15cm \noindent Mortlock D.J., et al., 2011, Nature 474, 616.

\vskip0.15cm \noindent Rhook K.J., Haehnelt M.G., 2008, MNRAS 389, 270.

\vskip0.15cm \noindent Richards G., et al. 2006, ApJSS, 166, 470

\vskip0.15cm \noindent Rosati P., et al., 2009, STECF 46, 12.

\vskip0.15cm \noindent Schmidt M., Schneider D.P., Gunn J.E., 1995, AJ 110, 68.

\vskip0.15cm \noindent Shankar F., Crocce M., Miralda-Escude J., Fosalba P., Weinberg D.W., 2010, ApJ, 718, 231.

\vskip0.15cm \noindent Shang Z., Wills B.J., Wills D., Brotherton M.S., 2007, AJ, 134, 294

\vskip0.15cm \noindent Sulentic J.W., Stirpe G.M., Marziani P., Zamanov R., Calvani M., Braito V., 2004, A\&A 423, 121.

\vskip0.15cm \noindent Telfer R.C., Zheng W., Kriss G.A., Davidsen A.F., 2002,
ApJ 565, 773.

\vskip0.15cm \noindent Trakhtenbrot B., Netzer H., Lira P., Shemmer O., 2011, ApJ, 730, 7.

\vskip0.15cm \noindent Tozzi P., et al., 2006, A\&A, 451, 457.

\vskip0.15cm \noindent Vanden Berk D.E. et al., 2001, AJ, 122, 549

\vskip0.15cm \noindent Wang Y. et al., 2010, MNRAS 439, 737.

\vskip0.15cm \noindent Willott C., et al. 2010, AJ 139, 906.

\vskip0.15cm \noindent Willott C. et al. 2010b, AJ 140, 546.

\vskip0.15cm \noindent Willott C., 2011, Nature 474, 583. 
\end{document}